# Corkscrew motion of *Trypanosoma brucei* is driven by helical beating of the flagellum and facilitated by its bent shape

**Authors:** Sizhe Cheng[1], Devadyouti Das[1], Mykhaylo Barchuk[1], Raveen Armstrong[2], Michele M. Klingbeil[2], Becca Thomases[3], Shuang Zhou[1,*]

**Affiliations:**

[1.] Department of Physics, University of Massachusetts, Amherst; Amherst, MA 01003, USA.

[2.] Department of Microbiology, University of Massachusetts, Amherst; Amherst, MA 01003, USA.

[3.] Department of Mathematical Sciences, Smith College; Northampton, MA 01063, USA.

[*]Corresponding author.

**Email:** zhou@physics.umass.edu.

**Author contributions:**

Conceptualization: SC, SZ

Methodology: SC, DD, MB, RA, MK, BT, SZ

Investigation: SC, DD, MB, RA, MK, BT, SZ

Project administration: SZ

Supervision: MK, BT, SZ

Writing – original draft: SC, SZ

Writing – review & editing: SC, MK, BT, SZ

**Competing interests:** Authors declare that they have no competing interests.

**Preprint servers**: arXiv:2512.10021. arXiv.org perpetual, non-exclusive license 1.0

**Classification:** Physical Sciences / Biophysics and Computational Biology

**This PDF file includes:**

Main Text
Figures 1 to 5

**Abstract**

In the pathogenic parasite *Trypanosoma brucei*, a laterally attached flagellum drives rapid deformation of the complex cell body, producing puzzling dynamics. High-speed defocusing imaging reveals that surface points trace flower-like patterns in transverse planes. The petals arise from clockwise flagellar beating, which generates a right-handed helical wave propagating from the anterior tip along the body, advancing the cell like a twisted corkscrew. The central lobes result from slower counterclockwise body rotation required to balance the active torque. The bent cell shape underneath the flagellum superimposes these two chiral motions at different radial distances, producing the observed patterns. Three-dimensional hydrodynamic simulations using the method of regularized Stokeslets reproduce these dynamics and show that bent cell shape enhances swimming, suggesting an adaptive advantage of *T. brucei*'s morphology.

**Significance Statement**

*Trypanosoma brucei* is a single-cell parasite causing sleeping sickness across 37 sub-Saharan countries, threatening ~60 million people. Revealing its locomotion mechanism is critical to understand its lifecycle in fly vectors and mammalian hosts; however, progress has been hindered by the rapid and complex cell deformation driven by a laterally attached flagellum. By tracking the 3D motion of fluorescent particles attached to cell surface, we demonstrate that tip-to-base helical waves generate self-propulsion like an active corkscrew, with reactive torque inducing counter-rotation of the cell body. These findings inform future investigations of the underlying biomechanics, suggest potential routes for therapeutic intervention, and inspire biomimetic micro-robotic designs. The defocused particle tracking technique is broadly applicable for deciphering the dynamics of other complex microswimmers.

**Main Text**

**Introduction**

Eukaryotic unicellular microswimmers exhibit a great variety and flexibility of swimming gaits to adapt to their natural habitats (1). For example, *Chlamydomonas reinhardtii* uses synchronized "breaststrokes" of the flagella pair (2) to self-propel in bulk fluid, but can beat asynchronously to push off the boundary (3) or steer towards light (4); spermatozoa cells propagate planar bending waves along their "tail" in regular medium (5), but modulate its curvature (6) or change it to helical waves (7, 8) in response to high-viscosity fluids; parasites *Crithidia* spp. (9) and *Leishmania* spp. (10) propagate uniform tip-to-base bending waves on the front-mounted single flagellum during forward motion, but initiate a short burst of high-curvature base-to-tip waves at any point of the flagellum to reorient the body (11). The core structure enabling these diverse behaviors is a highly unified microtubule cytoskeleton called axoneme (12), an organelle emerged early in evolution (13) and shared by all eukaryotes, including multicellular organisms (14), to transport fluids. By orchestrating thousands of dynein motors distributed along its entire length, axoneme generates rich forms of deformation waves to best fulfil its biological purpose, and respond to the sensory signals induced by changed environment (12, 14). Understanding the interplay between the swimming mechanism, biological function, and the environmental cues provides valuable insights into the biomolecular coordination underlying axoneme motion (15), the mechanics of cellular structures (10, 16), the influence of fluid property (17), and the ecology of the living system (18). Furthermore, swimming mechanisms discovered in microorganisms often inspire the design of artificial systems capable of directed and purposeful movement (19, 20).

In contrast to some well-understood microswimmers, the swimming mechanism of parasite *Trypanosoma brucei* remains elusive (21). *T. brucei* is the causative pathogen of sleeping sickness, which threatens the health and economic development of nearly 60 million people in 37 sub-Saharan African countries (22). To complete its dixenous lifecycle between insect and mammalian hosts, *T. brucei* must overcome tremendous physical challenges (23, 24), including travelling back and forth in the alimentary track of tsetse fly at different development stages (25), burying themselves into the peritrophic matrix of insect gut (26), navigating through the blood vessels of the mammals (27, 28), penetrating their central nervous system barrier to reach the brain (29), and using hydrodynamic drag to remove antibody attacks (30). It remains a long-standing puzzle how they manage to self-propel with a laterally attached flagellum and accomplish such demanding tasks. Although their name, derived from the Greek *trypanon* (τρύπανον) meaning "auger" and *soma* (σῶμα) meaning "body", suggests a corkscrew-like motion (31), observations based on traditional microscopy failed to provide concrete evidence and fueled the debate between two prevailing models. One model (32) suggests that the flagellum generates bihelical waves of alternating chirality, creating "kink waves" to propel the cell, and resulting in oscillatory rotation of the body around the swimming direction. Another model, however, claims that flagellum beats in a planar fashion to self-propel like many other eukaryotes, but the beating plane rotates and creates a left-handed cell shape that spontaneously and continuously rotates as it translates (28, 33–35). A major source of this puzzle is that conventional high-speed video data only shows a two-dimensional (2D) projection of a transparent *T. brucei* undergoing three-dimensional (3D) dynamics. In well-established structure-from-motion studies, a rotating transparent cylinder is a classic example where opposite-direction rotations produce identical 2D motion patterns (36, 37). Therefore, to accurately reconstruct the motility dynamics, a new method capable of revealing 3D information is crucial (38).

In solving this puzzle, we establish a new model based on the analysis of 3D dynamics of cell body points recorded at high spatiotemporal resolution. Different from previous models, we confirm that the body of *T. brucei* performs a high-frequency chiral beating in the *clockwise (cw)* direction (viewed from the posterior end), propagating a right-handed helical wave, which travels down the cell body and generates thrust force like a twisted right-handed corkscrew. Meanwhile, the reactive torque, required by the torque-free constraint, causes the entire cell body to precess in the *counterclockwise (ccw)* direction around the average swimming axis ( $\boldsymbol{e}_z$ ). Interestingly, this more obvious, slower counterrotation, coupled with forward motion, is likely what created the impression of corkscrew motion in early observations and inspired the organism's name. On the contrary, the high-frequency chiral beating that actually drives self-propulsion is far more subtle and can only be clearly resolved through high-speed 3D imaging. The bent cell shape and the lateral attachment of the flagellum couple these two chiral motions at different radial distances from $\boldsymbol{e}_z$ , resulting in the characteristic flower-like trajectories of the labelling particles on the transverse plane. Numerical simulation confirms this model by reproducing closely matched translational and rotational velocities, surface-point trajectories, Fourier signatures, and 2D images of the cells. It further demonstrates that a gently bent cell shape facilitates translation, which might explain the adaptation of such cell morphology.

## Results

### 3D Motion of *T. brucei* body revealed by attached fluorescent particles

The procyclic form of *T. brucei* presented in this study is a highly motile stage of the insect cycle, featuring an asymmetric slender body of $\sim 20$ μm in length, with the thickest part, of $\sim 3$ μm in diameter, located close to the posterior end, roughly corresponding to the location of nucleus and the flagellum pocket. From this site, a single flagellum emerges and extends towards the anterior end, while remaining laterally attached on the cell surface along most of its length, with only a small segment free at the anterior end (21). We labelled the surface of 22 *T. brucei* cells with 1-2 fluorescent particles per cell at random locations and tracked their 3D dynamics (Fig. 1) using defocusing particle tracking microscopy (39, 40) (DPTM; Materials and Methods). Recorded at sub-micrometer spatial resolution and 100 frame per second rate, all labelled points exhibit periodic 3D motion (Fig. 1, 3D Trajectory). When projected on the transverse plane $\perp \boldsymbol{e}_z$ (the average swimming direction), the trajectories exhibit rich features that vary with the labelling position. While the flagellar pocket traces simple *ccw* loops with a large radius of ~ 2 μm (Fig. 1D), other body locations trace superimposed shapes of small *cw* loops ("petals") on a larger *ccw* circle ("central lob"), forming flower-like patterns (Fig. 1A-C, E). As the distance from the flagellum pocket increases, the size of these "petals" increases in both radial ( $\boldsymbol{e}_r$ ) and azimuthal ( $\boldsymbol{e}_\theta$ ) directions (Fig. 1A-C). Interestingly, flower-like patterns even appear at the flagellum-absent posterior end (Fig. 1E). The number of petals per large loop (*n*) remains roughly constant over time for a given cell, but varies between cells (Fig. 1).

We apply temporal analysis on the 22 labelled cells to separate their translation (linear velocity $\boldsymbol{U}$ ), rotation (angular velocity $\boldsymbol{\Omega}$ ), and the internal motions generated by flagellum beating, a.k.a. the "swimming gait" $\boldsymbol{G}(\boldsymbol{r}_s,t) \equiv \dot{\boldsymbol{r}}_s \equiv \partial \boldsymbol{r}_s / \partial t$ . For a given surface point marked by $\boldsymbol{r}_s$ in the body-fixed frame, its instantaneous velocity $\boldsymbol{u}(\boldsymbol{r}_s,t)$ is the superposition of these three modes: $\boldsymbol{u} = \boldsymbol{G} + \boldsymbol{U} + \boldsymbol{\Omega} \times \boldsymbol{r}_s$ . For all 22 points, the time-dependent position functions in the cylindrical coordinates, $z(t)$, $\theta(t)$ and $r(t)$, can be decomposed into linear and oscillatory functions (Fig. 2A, B). Fitting $z(t) = Ut$ yields linear translation at $U = 4.3 \pm 1.7$ μm·s$^{-1}$; fitting $\theta(t) = \Omega t$ and $r(t) = r_0$ reveals *ccw* rotations with $\Omega = -7.4 \pm 3.4$ rad·s$^{-1}$ at a radial distance $r_0 = 2.0 \pm 0.4$ μm . Fourier analysis on $r(t) - r_0$ and $\theta(t) - \Omega t$ reveals a common frequency of $f = 6.9 \pm 2.2$ Hz, with $|2\pi f/\Omega| \approx 6.8 \pm 2.7 \approx n$ matching the number of "petals per flower", suggesting that this Fourier peak corresponds to the high-frequency *cw* motion. Interestingly, the same Fourier peak is also found on $z(t) - Ut$ , indicating that the oscillation has an axial component. The peak heights in the Fourier spectrum correspond to $A_r$ , $A_\theta / r_0$ , and $A_z$ , where $A_r$ , $A_\theta$ , $A_z$ are oscillation amplitudes along $\boldsymbol{e}_r$ , $\boldsymbol{e}_\theta$ , and $\boldsymbol{e}_z$ , respectively. Across the cell body, $r_0$ remains nearly constant (Fig. 2E), while $A_r$ , $A_\theta$ , and $A_z$ show a minimal at the flagellum pocket (Fig. 2F).

Overall, the motion of body points clearly indicates that *T. brucei* beats its flagellum in a chiral, *cw* motion at high frequency, generating a rapid right-handed helical wave from its anterior end, which can propagate beyond the flagellum pocket and reach the posterior end. Meanwhile, the cell body rotates more slowly in the *ccw* direction. Flower-like patterns traced by surface points resemble hypotrochoids, which can be generated by tracking a point on a spinning cylindrical surface whose axis simultaneously precesses around a different direction. Here the spinning corresponds to the chiral beating of the flagellum, and precession corresponds to the slow counterrotation of the cell body (Fig. S10 and SI Sec. 13). To quantitatively understand how such chiral beating leads to self-propulsion and this *ccw* rotation, we simulate the hydrodynamic outcomes using the method of regularized Stokeslets (41, 42).

## Numerical simulation based on helical beating and a bent body

Based on geometric parameters measured from 50 cells, we outline *T. brucei* body as three frustoconical sections, with a total axial length of $L = 20$ μm (Fig. 3A). The slender anterior section (along $\boldsymbol{e}_1$) has a length $L_1 = 7$ μm and a radius that increases linearly from 0.4 to 0.83 μm. The middle section ( $\boldsymbol{e}_2$, $L_2 = 9$ μm ) continues to expand and ends with a maximum radius of 1.5 μm at the flagellum pocket. From there, the posterior section ( $\boldsymbol{e}_3$, $L_3 = 4$ μm ) tapers back down to a radius of 0.4 μm . The three sections are joined with two bending angles $\alpha = (\boldsymbol{e}_1, \boldsymbol{e}_2)$ and $\beta = (\boldsymbol{e}_2, \boldsymbol{e}_3)$. The bent cell shape defines a body plane ( $\boldsymbol{e}_n$ ). For any beating symmetric about this plane, the swimming direction remains confined to the body plane, $\boldsymbol{e}_z \perp \boldsymbol{e}_n$ , and is generally misaligned with local body axis, $\boldsymbol{e}_z \neq \boldsymbol{e}_{1,2,3}$ (Fig. 3A; SI Sec. 7). Note that this minimal model ignores minor details such as the short free segment of flagellum to capture the major features of the cell shape (SI Sec. 11).

We discretize the cell surface into regularized point forces (Stokeslets) and describe the anterior-to-posterior helical wave as a superposition of two orthogonal transverse waves along local axis $\boldsymbol{e}_{1,2,3}$ , with a $\pi/2$ phase shift between them. One wave is perpendicular to the body plane, $A_n(s)\sin(\omega t - ks)\boldsymbol{e}_n$ . The other wave stays in the body plane, $A_p(s)\cos(\omega t - ks)\boldsymbol{e}_p$ , $\boldsymbol{e}_p \perp \boldsymbol{e}_{1,2,3}, \boldsymbol{e}_n$ , generating $\boldsymbol{e}_z$ oscillations unless $\alpha = \beta = 0$ (SI Sec. 12 and Fig. S8). Here, *s* is the axial distance measured from the anterior end, $\omega = 2\pi f$ the angular frequency, $k = 2\pi / \lambda$ the wave number, and $\lambda = 10$ μm the wavelength of the helical wave (SI Sec. 8). Beating amplitudes $A_n(s)$ and $A_p(s)$ are specified at 4 locations and linearly interpolated in between (Fig. 2F). At the anterior end ( $s = 0$ ), we independently choose $A_n$ and $A_p$ in the $(0.5 - 2.8)$ μm range to capture the variability across cells. At the first junction ( $s = 7$ μm ), we set $A_n, A_p = (0.4 - 1.0)$ μm , and reduce both to 0 at the flagellum pocket ( $s = 16$ μm ). At the posterior end ( $s = 20$ μm ), we set $A_n, A_p = (0.2 - 0.8)$ μm , due to the consistently observed flower-like patterns there. Longitudinal waves are excluded, since the eukaryotic flagellum is inextensible.

With the *T. brucei*'s swimming gait defined in the model, we numerically simulate the hydrodynamics using the method of regularized Stokeslets (41, 42) with free-swimming constraints (SI Sec. 8). Using a set of computational parameters $(\alpha, \beta, A_n(s), A_p(s))$ (SI Table S1) based on the biological ones determined in experiments, the simulation results closely match experimentally measured *U* and $\Omega$ , while reproducing similar trajectories (Fig. 1, Simulation), position functions (Fig. 2C), and Fourier peaks (Fig. 2D) of corresponding body points. Meanwhile, the 2D projection of the modeled swimmer matches the phase contrast microscope images frame-by-frame for seconds (Fig. 4A, B, SI Movie S2). Over one beating period of $\sim 150$ ms , the view of transverse plane $(\boldsymbol{e}_r, \boldsymbol{e}_\theta)$ (Fig. 4C) and the time-lapse 3D shape at various angles (Fig. 4D-F) show that as the right-handed helical wave travels down the body, the anterior section rotates in *cw* direction while the whole body rotates in *ccw* direction.

To understand how body shape and beating amplitudes affect the motility of *T. brucei*, we independently vary $\alpha$ in the $0 - 60°$ range in $15°$ increment and $A_n(0)$ and $A_p(0)$ in the $1 - 3$ μm range in 1 μm increments, while fixing other parameters: $\beta = 15°$ , $A_n = A_p = 0.8$ μm at first junction, 0 μm at flagellum pocket, and 0.6 μm at posterior end (Fig. 2F, SI Table S2). The time-independence of Stokes equation requires $U \propto f$ and $\Omega \propto f$ , and inspires us to characterize the motility of *T. brucei* using dimensionless translational velocity $k_v = \frac{U}{\lambda f}$ (ratio of swimming to wave speed), and rotational velocity $k_\Omega = \frac{|\Omega|}{2\pi f}$ (ratio of the body rotation to the flagellum beating rate). In this motility space, simulated motilities are presented as polygons connecting discrete simulation points with the same $\alpha$ but different amplitude profiles $(A_n, A_p)$ , or as dashed lines connecting simulation points with the same $(A_n, A_p)$ but different $\alpha$ (Fig. 5). In each polygon, as either $A_n(0)$ or $A_p(0)$ increases, both $k_v$ and $k_\Omega$ increase; interchanging $A_n(0)$ and $A_p(0)$ values does not affect motility significantly. Increasing $\alpha$ generally shift the polygons towards the origin. However, for a fixed $\left(A_n(0), A_p(0)\right)$, especially at large values, the dashed lines trace peculiar arcs in the motility space, with $k_v$ maximizes at $\alpha$ around $30-45^\circ$ (Fig. 5 inset). This angle matches well with the bending angle measured in surface-attached, non-swimming *T. brucei* (Fig. S6).

The experimental results can be compared with the simulations in the motility space ($k_\Omega$ vs. $k_v$). They are quantified from direct measurements of $U$, $\Omega$ and $f$, using an estimated wavelength $\lambda = 10$ μm consistent with values reported in previous studies (10, 21, 24, 32, 33, 35). However, since these swimmers didn't attach to substrate and each carries only 1 or 2 fluorescent particles, we cannot measure $\alpha$, $\beta$, or the entire beating profile $\left(A_n, A_p\right)$. Although one-to-one comparison is not possible, the scattered experimental points overlap with the region predicted by the simulations and show a generally positive correlation between $k_v$ and $k_\Omega$ (Fig. 5).

## Discussion

### 3D measurements resolve ambiguities inherent in 2D projections

High-speed 2D imaging has been widely used to characterize the swimming dynamics of microorganisms. However, reconstructing 3D motion from a 2D view is inherently ambiguous, especially for nearly transparent swimmers where the projected image does not clearly separate the near and far sides of the body (36–38). This ambiguity has led to different interpretations of the same swimming behavior. Built on high-speed DIC observations of *T. brucei*, the bihelical model believes that the forward cell motion ($v = 5 \pm 2$ μm·s$^{-1}$) is driven by traveling kinks (backward at $v_k = 85 \pm 18$ μm·s$^{-1}$) which separate chiral waves of opposite handedness, similar to *Spiroplasma*. The 180° flips of the anterior at $f_{ant} = 19 \pm 3$ Hz and of the posterior at $f_{post} = 5 \pm 3$ Hz deny continuous body rotation, and imply a puzzling frequency gradient along the cell body (32). In contrast, based on bright-field tomographic reconstructions (28, 43) and multifocal plane fluorescence microscopy (44), the plane-rotational model (28, 33–35) suggests that flagellum beating at $f_{pb} = 18.3 \pm 2.5$ Hz is locally planar and responsible for forward motion at $v = 5.7 \pm 0.11$ μm·s$^{-1}$. But the beating plane rotates with the laterally attached flagellum, deforming the cell body into a left-handed helix. The continuous *ccw* body rotation at $f_r = 2.8 \pm 0.4$ Hz, corresponding to roughly 6 beats per turn, is a passive effect due to the left-handed helical body undergoing translation. Other studies focus on the statistical features of swimming motion and identify similar temporal characteristics. For example, by analyzing cell motions in 2D, Zaburdaev et al. identify a time scales of $\tau \approx 0.1$ s capturing the rapid, micron-scale oscillations of the trajectory (45).

Our direct 3D measurements of surface point motion eliminate artifacts arising from 2D projections. The motion reveals a unidirectional chiral beating of the flagellum, rather than planar beating, accompanied by a continuous *ccw* rotation of the cell body instead of an oscillatory one. The characteristic time scales of beating and rotation remain comparable to those reported previously. The chiral beating frequency, $f = 6.9 \pm 2.2$ Hz, corresponds well to the $f_{ant}/2$ in bihelical model, $f_{bp}/2$ in plane-rotational model, and $1/\tau$ in the trajectory analysis. The body rotation frequency, $f_{ccw} = \left|\Omega/2\pi\right| = 1.2 \pm 0.5$ Hz, is close to $f_{post}/2$ and $f_r$ in corresponding models. The lower values are consistent with the slower swimming velocity $U = 4.3 \pm 1.7$ μm·s$^{-1}$ in our experiments, which can be attributed to biological variations.

**Active torque drives the helical beating. Reactive torque drives the counterrotation.**

In our simulation, it is hard to directly quantify the active torque $\boldsymbol{T}_{cw}$ enabling the high-frequency chiral beating, since it is distributed across the entire cell body. However, the resultant thrust force $\boldsymbol{F}_{th}$, which drives forward motion, and the reactive torque $\boldsymbol{T}_{ccw} = -\boldsymbol{T}_{cw}$, which drives counterrotation of the body frame (Fig. 4C–F and SI Movie S3), can be quantified in a set of complementary simulations where a swimmer is fixed in position and orientation ($\boldsymbol{U} = 0$, $\boldsymbol{\Omega} = 0$, SI Sec. 14), but allowed to beat normally. In these cases, the net force and torque obtained by summing over all surface points correspond to the holding force $-\boldsymbol{F}_{th}$ and torque $-\boldsymbol{T}_{ccw}$ due to the body-fixed constraints. By averaging simulations results over one beating period, we obtain $\boldsymbol{F}_{th}^{b}$ and $\boldsymbol{T}_{ccw}^{b}$ which generally misalign with $\boldsymbol{e}_z$, the free-swimming direction. $\boldsymbol{F}_{th}^{b} \cdot \boldsymbol{e}_z > 0$ and $\boldsymbol{T}_{ccw}^{b} \cdot \boldsymbol{e}_z < 0$ are consistent with their expected direction according to the translation and rotation of the cell.

The holding force and torque obtained in body-fixed simulations equal the hydrodynamic drag force $\boldsymbol{F}_{drag} = -\boldsymbol{F}_{th}$ and torque $\boldsymbol{T}_{drag} = -\boldsymbol{T}_{ccw}$ in the free-swimming cases. At low Reynolds number, $\boldsymbol{F}_{drag}$ and $\boldsymbol{T}_{drag}$ linearly depend on $\boldsymbol{U}$ and $\boldsymbol{\Omega}$ through the resistance matrix, which allows us to identify their mechanical origin through decomposition:

$$-\begin{pmatrix} \boldsymbol{F}_{th} \\ \boldsymbol{T}_{ccw} \end{pmatrix} = \begin{pmatrix} \boldsymbol{F}_{drag} \\ \boldsymbol{T}_{drag} \end{pmatrix} = \begin{pmatrix} \boldsymbol{F}_{t} + \boldsymbol{F}_{r} \\ \boldsymbol{T}_{t} + \boldsymbol{T}_{r} \end{pmatrix} = \begin{pmatrix} \overline{\overline{A}} & \overline{\overline{B}} \\ \overline{\overline{B}} & \overline{\overline{D}} \end{pmatrix} \begin{pmatrix} \boldsymbol{U} \\ \boldsymbol{\Omega} \end{pmatrix}$$

Here, $\boldsymbol{F}_t = \overline{\overline{A}}\boldsymbol{U}$ ($\boldsymbol{T}_t = \overline{\overline{B}}\boldsymbol{U}$) and $\boldsymbol{F}_r = \overline{\overline{B}}\Omega$ ($\boldsymbol{T}_r = \overline{\overline{D}}\Omega$) are the drag forces (torques) due to cell translation and rotation, correspondingly. The matrix components, $\overline{\overline{A}}$, $\overline{\overline{B}}$, and $\overline{\overline{D}}$, are entirely determined by the fluid viscosity, size, and instantaneous shape of the swimmer. Although their full tensorial forms can describe instantaneous force and torque balance with arbitrary $\boldsymbol{U}$ and $\boldsymbol{\Omega}$, we focus on the long-time limit, with $\boldsymbol{e}_x$ and $\boldsymbol{e}_y$ terms averaging to zero and only $\boldsymbol{e}_z$ terms survive. With $\boldsymbol{U}^{\infty} = U\boldsymbol{e}_z$ and $\boldsymbol{\Omega}^{\infty} = \Omega\boldsymbol{e}_z$, we can reduce $\overline{\overline{A}}$, $\overline{\overline{B}}$, and $\overline{\overline{D}}$ to scalars (see an example in SI Sec. 15):

$$-\begin{pmatrix} \boldsymbol{F}_{th}^{b} \\ \boldsymbol{T}_{ccw}^{b} \end{pmatrix} \cdot \boldsymbol{e}_z = \begin{pmatrix} F_{drag}^{\infty} \\ T_{drag}^{\infty} \end{pmatrix} \boldsymbol{e}_z = \begin{pmatrix} F_{t}^{\infty} + F_{r}^{\infty} \\ T_{t}^{\infty} + T_{r}^{\infty} \end{pmatrix} \boldsymbol{e}_z = \begin{pmatrix} A & B \\ B & D \end{pmatrix} \begin{pmatrix} U \\ \Omega \end{pmatrix} \boldsymbol{e}_z$$

We quantify the long-term resistance forces and torques and the resistance matrix by another set of complementary simulations (SI Sec. 15), where the swimmer, “frozen” in various beating phases as a rigid body, is translated along or rotated around $\boldsymbol{e}_z$. By setting $\Omega = 0$ ($U = 0$), summing over all force points on the surface, and averaging their z-components over one beating period, we obtain the external force $-F_t^{\infty} = -AU$ ($-F_r^{\infty} = -B\Omega$) and torque $-T_t^{\infty} = -BU$ ($-T_r^{\infty} = -D\Omega$), respectively. The values, $F_t^{\infty} + F_r^{\infty}$ vs $-\boldsymbol{F}_{th}^{b} \cdot \boldsymbol{e}_z$, and $T_t^{\infty} + T_r^{\infty}$ vs $-\boldsymbol{T}_{ccw}^{b} \cdot \boldsymbol{e}_z$, agree with each other within 3% and 4% of error, respectively, suggesting a nearly perfect cancellation of $\boldsymbol{e}_x$ and $\boldsymbol{e}_y$ terms (SI Sec. 16).

To investigate whether the chiral shape is enough to induce spontaneous body-frame rotation at the speed observed in experiments, we perform the complementary simulations with fixed $A_n(0) = A_p(0) = 3$ μm at the anterior end and varying $\alpha = 0° - 60°$ (SI Table S2). We find that $T_t^\infty = \langle \boldsymbol{T}_t \cdot \boldsymbol{e}_z \rangle > 0$ , $T_r^\infty = \langle \boldsymbol{T}_r \cdot \boldsymbol{e}_z \rangle > 0$ , and their ratio maximizes at $\alpha = 0°$ with $\left(T_t^\infty / T_r^\infty\right)_{\max} \approx 0.5\%$ (SI Sec. 15). This result shows that if the swimmer is only translated at $\boldsymbol{U}^\infty = U\boldsymbol{e}_z$ but not driven by the reactive torque $\boldsymbol{T}_{ccw}$ , the spontaneous rotation due to chiral shape is not only *cw* but also >200 times slower than experimental result, contradicting the plane-rotational model.

**Bent body shape for faster swimming**

Although some previous simulations (28, 33–35) modeled the static shape of *T. brucei* body as straight and axially symmetric, our study indicates that a bent shape is not only critical to produce experimental observations, but also important for faster swimming. Direct experimental support comes from phase contrast images of cells attached to glass substrate showing pronounced bent shapes (SI Sec. 7, Movie S4), agreeing with prior electron microscopy studies (21). Bent shape is also essential to produce the measured surface point dynamics by placing the *cw* beating and *ccw* rotation at different radial distances to create flower-like patterns, trajectory functions, and Fourier signatures. In the hypotrochoid analogy, deviating the spin axis from the precession axis allows the loops traced by spin to spread over a larger circle traced by precession, while aligning them only produces tightly packed spin circles. Indeed, our simulation on straight swimmer shows tightly packed loops, different $r(t)$ and $\theta(t)$ functions, and absence of $\boldsymbol{e}_z$ oscillations (SI Sec. 12, Fig. S8), conflicting with experiments.

Facilitation of locomotion, evident by the maximization of $k_v$ at $\alpha = 30° - 45°$ in simulations (Fig. 5 inset), can be a possible reason for adaptation of bent cell shape. We can understand this effect from how $\boldsymbol{F}_{th}^b$ , $\boldsymbol{T}_{ccw}^b$ , and resistance matrix changes in response of $\alpha$ . As $\alpha$ increases, the misalignment angle between the cell body and $\boldsymbol{e}_z$ increases (Fig. S11A), leading to a reduced $\boldsymbol{F}_{th}^b \cdot \boldsymbol{e}_z$ and $\left|\boldsymbol{T}_{ccw}^b \cdot \boldsymbol{e}_z\right|$ (Fig. S11C, D), which then requires less $\left|F_t^\infty\right| + \left|F_r^\infty\right|$ and $T_t^\infty + T_r^\infty$ to balance. Meanwhile, translational resistance $|A|$ remains nearly unchanged (<2%), rotational resistance $|D|$ increases significantly (>50%), and most interestingly, hydrodynamic chirality $B$ decreases drastically (>73%) (Fig. S11B). Combination of these changes causes the following effects as $\alpha$ increases. (i) The decrease of $T_t^\infty + T_r^\infty \approx T_r^\infty = |D||\Omega|$ and increase of $|D|$ lead to a strong reduction of $|\Omega|$ (Fig. 5 inset). (ii) Combined with decrease of $B$ , $\left|F_r^\infty\right| = B|\Omega|$ significantly reduces in the force balance (Fig. S11D): $\boldsymbol{F}_{th}^b \cdot \boldsymbol{e}_z = \left|F_t^\infty\right| + \left|F_r^\infty\right| = |A|U + B|\Omega|$ . (iii) As a result, even with reduced $\boldsymbol{F}_{th}^b \cdot \boldsymbol{e}_z$ , $\left|F_t^\infty\right| = |A|U$ peaks at $\alpha = 30° - 45°$ (Fig. S11D), which leads to maximized $U$ (Fig. 5 inset), since $|A|$ is practically constant. In other words, by increasing the rotational resistance, reducing hydrodynamic chirality, and maintaining a constant translational resistance, a bent cell body can significantly reduce the drag force due to body rotation to facilitate translation, resulting in maximized motility at an intermediate $\alpha$ value. A similar enhancement of propulsion by body curvature was reported on theoretical study of bacterial motility, which suggested that curved bodies may improve propulsion through a trade-off between translational and rotational resistance, reducing the power wasted by body counterrotation (46).

In summary, the discovery of this swimming gait establishes a new baseline for understanding *T. brucei* motility in its complex habitat. Although current work only focuses on their locomotion in the bulk of the fluid, the experimental and simulation techniques can be used to investigate their behavior near surface, in external flows, or under narrow confinements. With their basic swimming gait in Newtonian fluid established, future explorations can also be towards understanding their motion in non-Newtonian fluids (27, 28) and soft solids (26), which are critical parts of their dixenous infections in flies and in mammals. The underlying biomechanics (24), i.e., how *T. brucei* generates such helical beating, are beyond the scope of this work; a full understanding will require interdisciplinary investigations on the underlying biology (47–49), the mechanical roles of different components in the flagellum complex (50, 51), its coupling to the cell body (52–55), and the fluid mechanical effects at different settings (56, 57). Fully revealing the biomechanics of *T. brucei*'s swimming gait may open new avenue for mitigating sleeping sickness and inspire new designs of biomimicking microrobots.

## Materials and Methods

### Trypanosome Cell Maintenance

Wildtype Procyclic form *Trypanosoma brucei* strain Lister 427 cells are cultured at 27 °C in SDM-79 medium supplemented with 15% heat-inactivated fetal bovine serum. Cells are periodically diluted with fresh culture medium to maintain a density between $5\times10^{6}$ and $1\times10^{7}$ cells/ml.

### Cell chamber and fluorescent particle attachment

The tape chamber is assembled by bonding a standard glass slide and a 0.17 mm thick cover slip with a piece of 90 μm thick double-sided tape working as a spacer. The tape has a pre-punched 10 mm by 10 mm square hole in the middle to contain the cultural medium. We use 0.52 μm Fluoresbrite® YG Carboxylate microspheres to label *T. brucei* cells. Before imaging, a 10 μl drop of *T. brucei* suspension is placed at one side of the chamber, while another 10 μl drop of water suspension of fluorescent particles at a concentration of $1.82\times10^{10}$ particles/ml is placed at the opposite side. When the 2 drops make contact in the middle of the chamber, a gradient of growth media concentration is consequently established. When *T. brucei* swims through the contact region, fluorescent particles attach to *T. brucei* at random locations. While some cells become immotile after being exposed to low concentration of grow media, active cells tend to swim back to the side of regular growth medium concentration while carrying fluorescent particles with them. We track the 3D motion of the fluorescent particles when the cells are sufficiently far away from the contact line and chamber's walls to eliminate potential influences of the growth medium concentration change and the boundary effect. The strength of particle attachment is verified in optical tweezers experiment (SI Sec. 1) to ensure these fluorescent labels are not displaced during regular swimming motion.

### Dual-channel defocusing particle tracking microscope (DPTM)

We reconfigure a Nikon Eclipse Ti-E inverted microscope (Fig. S1A) to simultaneously imaging phase contrast channel and defocused fluorescent channel, using a Nikon S Plan Fluor ELWD 40× phase contrast objective (N.A. 0.6). The objective has a correction collar to compensate for or introduce spherical aberrations. *T. brucei* cells marked with fluorescent particles are illuminated by a transmitted light ( $\lambda = 675$ nm ), and excited by an epi-illumination of $\lambda = 445$ nm . After the objective, a beam splitter divides the light into the phase contrast and fluorescence channels (Fig. S1A). In the phase contrast channel, a long-pass filter HQ505lp (Chroma Technology) blocks the excitation light while allowing both the emission light ( $\lambda = 510$ nm ) and the transmitted light to pass through, resulting in images clearly showing the marked position. The fluorescence channel with the emission filter ( $\lambda = 510$ nm ) provides the defocused images of the particle marker. PCO.panda 4.2 and PCO.edge 4.2 cameras are used to capture images from the phase contrast and fluorescence channels with an exposure time of 10 ms at 100 fps, respectively (Movie S1).

In the traditional method of defocusing particle imaging, the particles above and below the focal plane show similar concentric rings, making the determination of absolute z-coordinate difficult (39, 58). We introduce additional spherical aberration to break this degeneracy. By intentionally setting the correction collar to 0.17 mm while using 1 mm thick glass substrate, we introduce finer concentric rings only when the particle is above the focal plane, allowing unambiguous $z$ determination (59). To obtain the z-coordinate of *T. brucei* markers, we compute the Pearson correlation coefficients $P_r$ between their defocused images and the pre-established calibration library, and look for the maximum of $P_r$ (SI Sec. 2). We interpolate $P_r$ between the library points to further improve the smoothness of the trajectory (SI Sec. 2). The uncertainty of our measurement mainly arises from the library step size ( $0.5\ \mu m$ ), resulting in a maximum uncertainty of $\pm 0.25\ \mu m$. The accuracy of this method was verified in the Brownian motion of free particles (SI Sec. 3).

**Acknowledgments**
This research is supported by NSF DMR 2239551 and Donal P. Reed Legacy Fund.

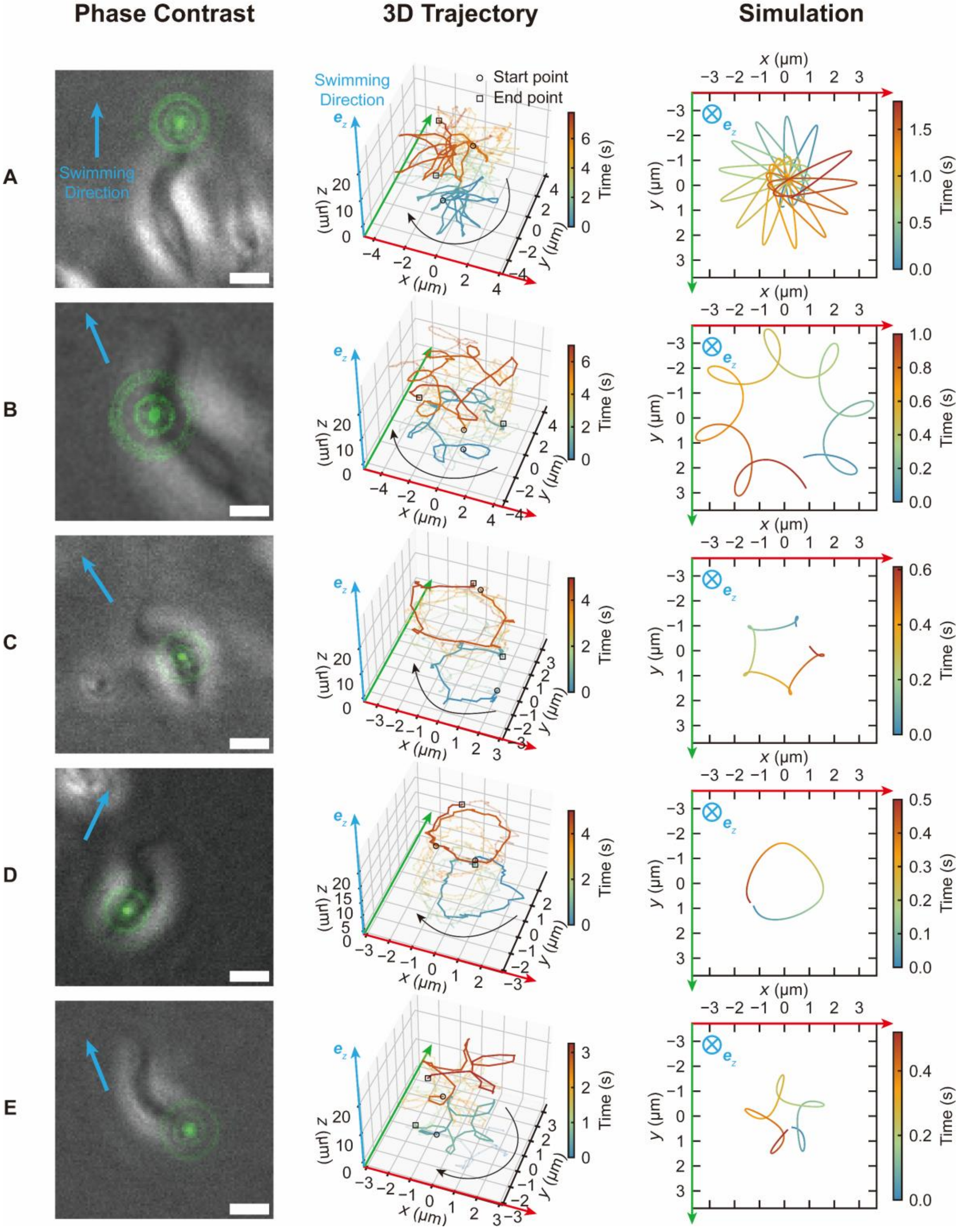

**Fig. 1. 3D trajectories of fluorescent particles at different locations on *T. brucei*'s body reveal chiral flagellum beating and body counterrotation. Phase contrast** images overlapped with ring patterns from DPTM show different labeling positions: (**A**) anterior end, (**B**) anterior cell body, (**C**) mid-cell body, (**D**) near the flagellar pocket, and (**E**) posterior end. Scale bar: 5 μm. The blue arrow indicates the swimming direction. **3D trajectories** of fluorescent particles reconstructed from DPTM images show periodic motion with flower-like patterns in transverse plane (A)-(E). The +z direction is the swimming direction, and the black curved arrow indicates the direction of body rotation. In the conventional posterior-to-anterior view, the overall pattern generated by body rotation is *ccw*, whereas the petals generated by chiral beating are *cw*. Note that the 3D trajectories are shown in an anterior-to-posterior view. Two full rotation periods are highlighted, with circles and squares indicating the start and end points of each highlighted segment, showing consistent repetition of the pattern over 3–6 s. **Simulation** trajectories of the corresponding labeled positions over one period of body rotation, projected onto the transverse plane, with the swimming direction pointing into the screen (posterior-to-anterior view), closely match the experiments. The parameters used in the simulations are listed in Table S1.

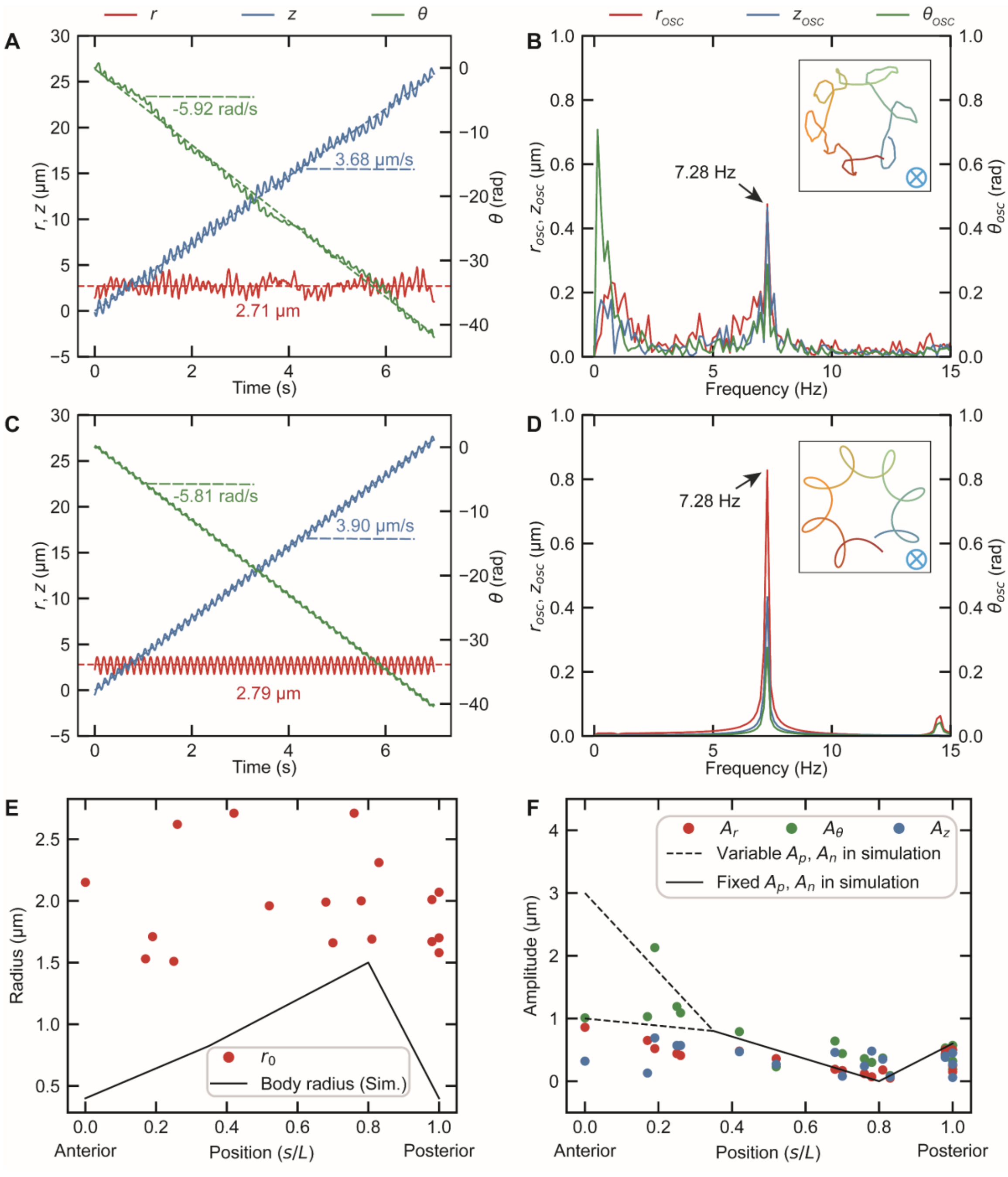

**Fig. 2. Dynamic parameters extracted from temporal analysis of surface points reveal detailed cell motion. (A)** Position functions of a particle attached at anterior section (Fig. 1B) in cylindrical coordinates show linear and oscillatory time dependences. Linear functions reveal a uniform translation accompanied by a slow *ccw* rotation. **(B)** Fourier spectra of the oscillatory components after removing the linear parts reveals an overlapping high-frequency peak for all three directions. **(C)** Simulation results show closely matched translational and rotational velocities, with **(D)** the same characteristic Fourier peaks. **(E)** Difference between the cell body radius (solid line, used in simulation) with the measured $r_0$ implies a bent cell shape. **(F)** Amplitude profiles $A_n$ and $A_p$ used in simulation (lines) are set according to $A_r$, $A_\theta$, and $A_z$ (points) measured from experiments. For simulation presented in Fig. 5, the amplitude profile of the anterior section falls between the dashed lines and is always linear with position.

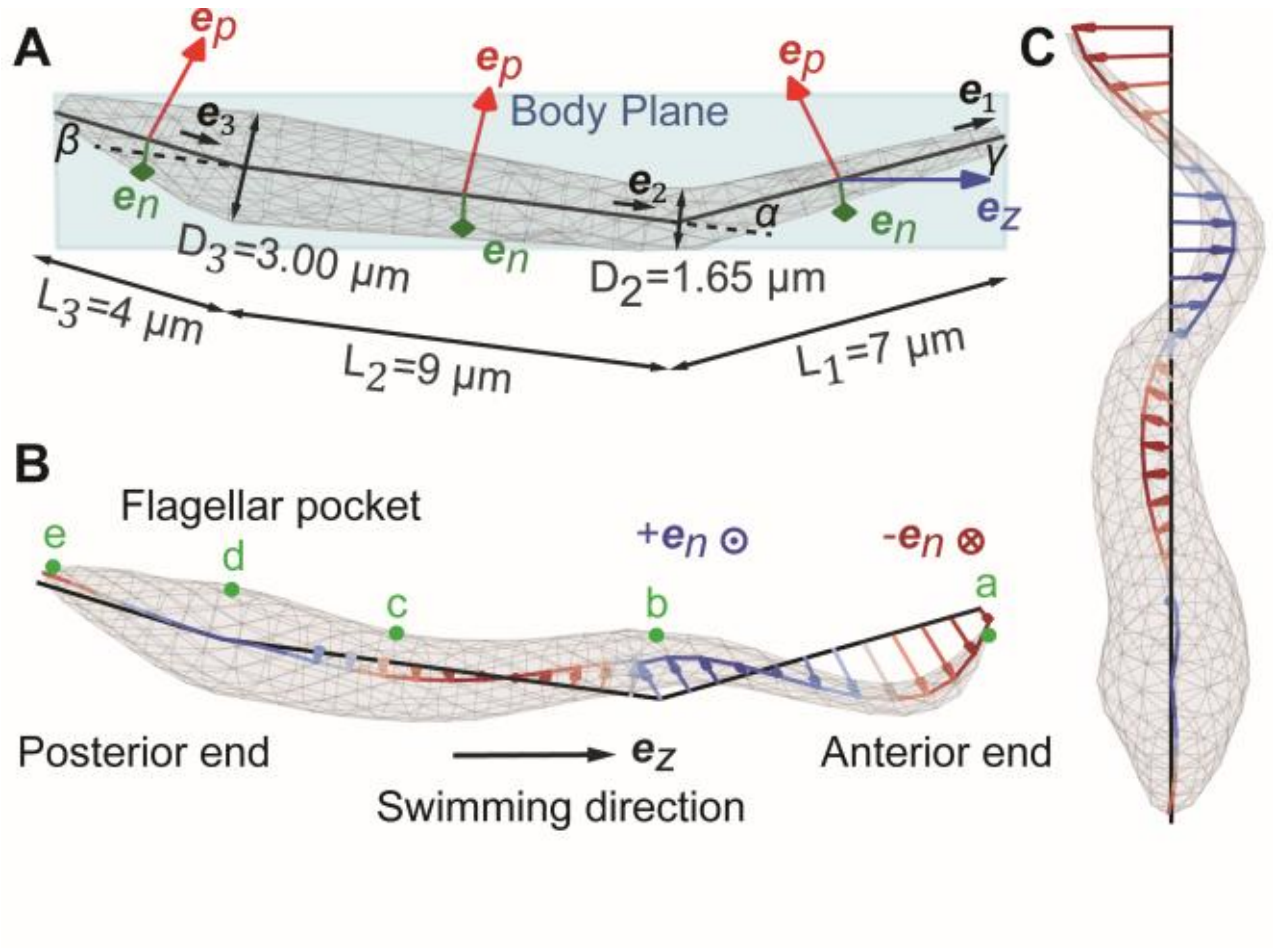

**Fig. 3. Body shape and swimming gait of *T. brucei* in simulation. (A)** Static body shape of *T. brucei* comprises of 3 frustoconical sections connected by 2 junctions, with their axes (black lines) defining the body plane. Two orthogonal, phase-shifted transverse waves along the body axes results in helical beating. $\boldsymbol{e}_n$ wave is orthogonal to the body plane. $\boldsymbol{e}_p$ wave stays in the cell body plane and is generally not orthogonal to the swimming direction $\boldsymbol{e}_z$. **(B)** Snapshot of a simulated *T. brucei* viewed normal to the body plane, showing the positions of the five fluorescent particles presented in Fig. 1. The gradient-colored arrows indicate the direction of the displacement vector of body axis with respect to the body plane: red points along -$\boldsymbol{e}_n$, and blue along $+\boldsymbol{e}_n$. **(C)** The same snapshot viewed along the body axes. Panels B and C together show a right-handed chiral wave during beating.

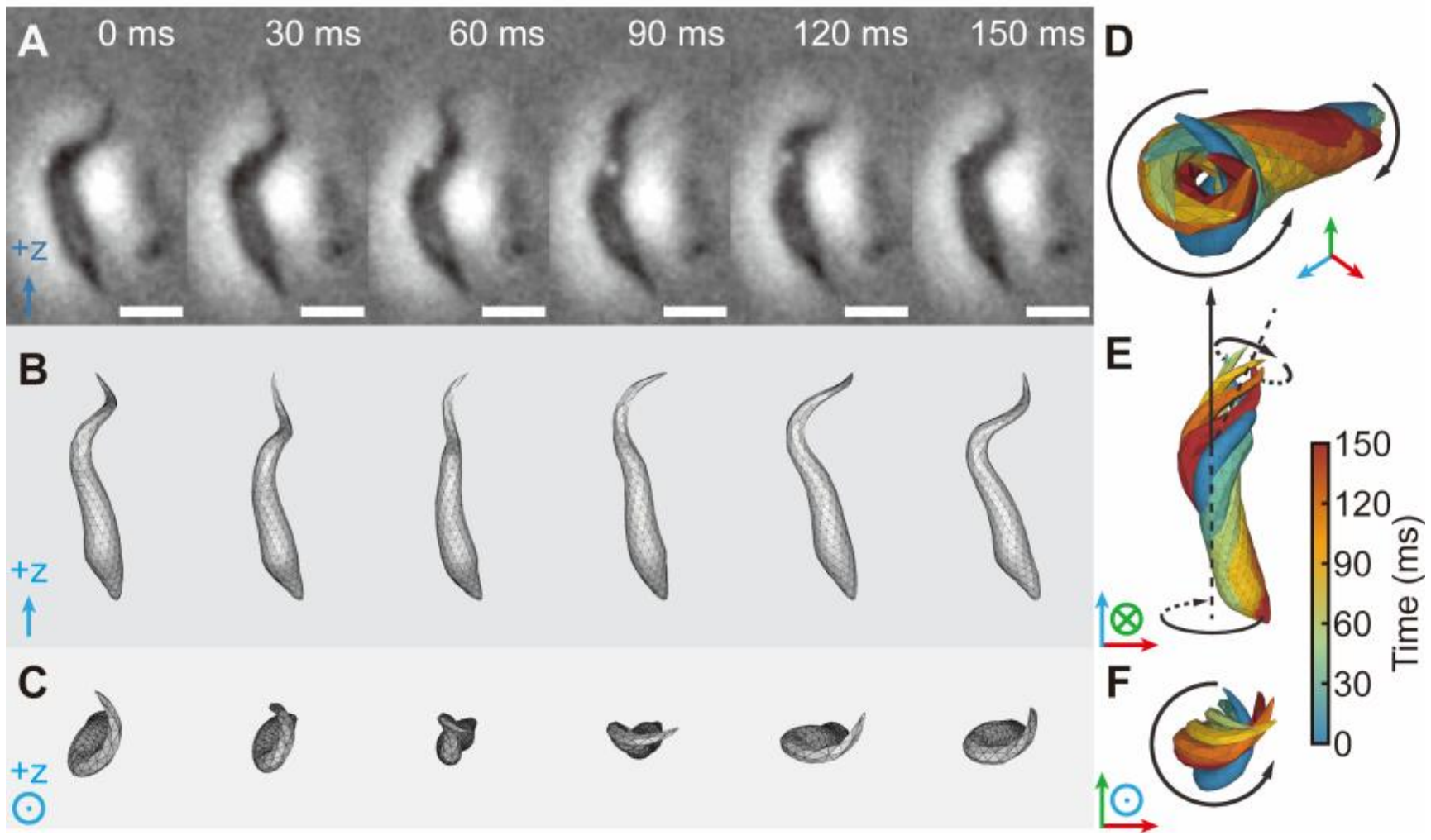

**Fig. 4. Dynamic cell shapes viewed from different angles reveal the helical beating gait. (A)** phase-contrast microscopy images acquired at 30 ms intervals over one beating period. The blue arrow indicates the swimming direction. Scale bar: 5 μm. **(B)** Simulation results shown from the same view as in (A), matching the experimental motion frame by frame. **(C)** Front view of the simulated motion at the same period reveals the corkscrew motion of the anterior part. **(D–F)** Time-lapse 3D view of the same period reveals the *ccw* rotation (view from posterior end) of the cell body while the anterior section corkscrewing in *cw* direction.

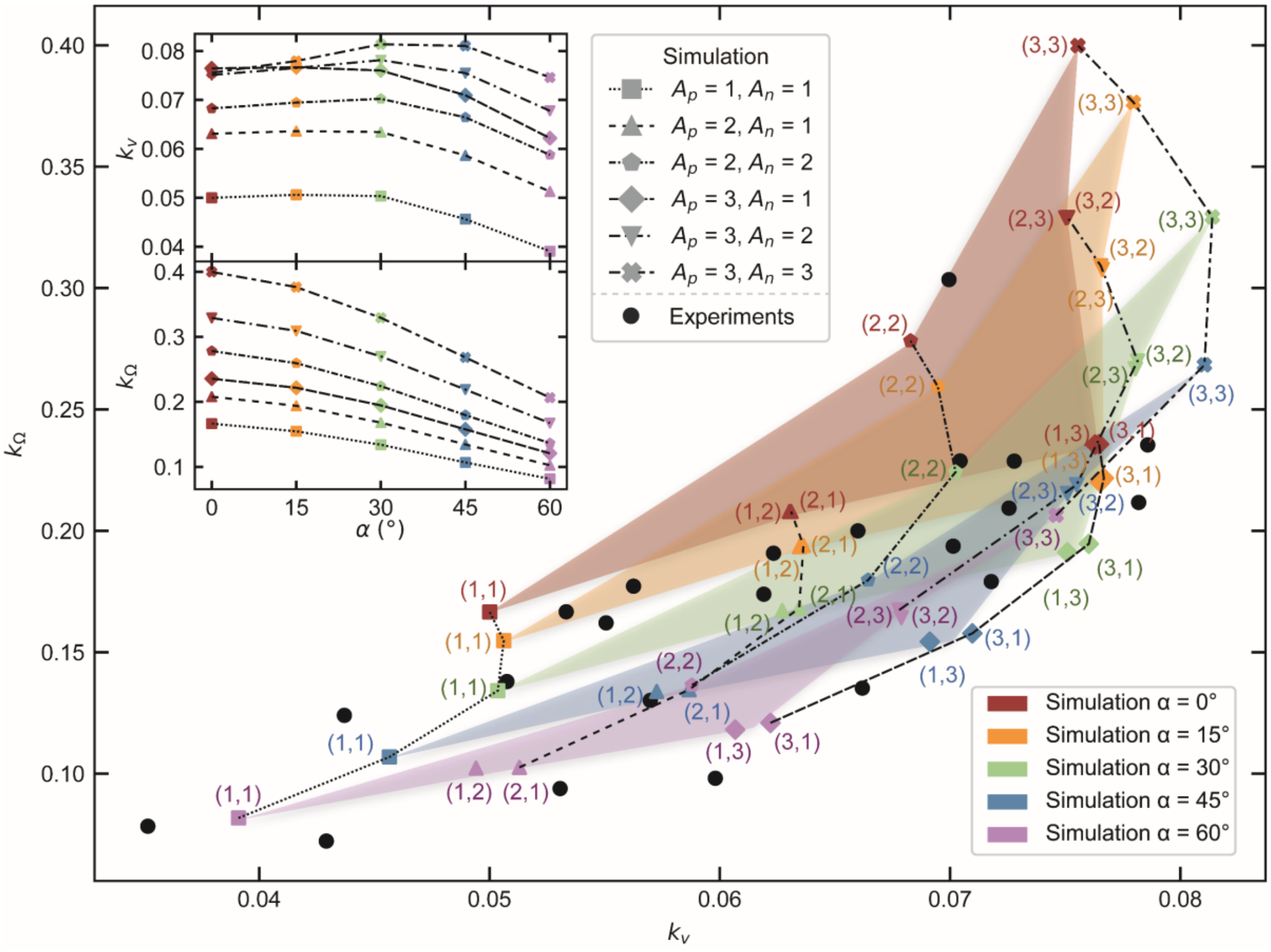

**Fig. 5. Experimental and simulated results in the motility space spanned by the dimensionless translational velocity $k_v$ and rotational velocity $k_\Omega$.** $k_v = \dfrac{U}{\lambda f}$ and $k_\Omega = \dfrac{|\Omega|}{2\pi f}$, where $U$, $\Omega$ and $f$ are measured from experiments or calculated from simulations, and $\lambda$ is fixed at an estimated value of 10 $\mu m$. **Black dots** represent experimental results, which show a generally positive relationship between $k_v$ and $k_\Omega$. **Colored symbols** of different shapes represent simulation results obtained under different parameter combinations. **The colored polygons** are formed by connecting simulation points at the same bending angle $\alpha$ but different beating amplitudes ( $A_p(0)$, $A_n(0) = 1$, 2, 3 $\mu m$ ), whereas **the dashed lines** connect simulation results with the same beating amplitudes but different $\alpha$, showing a non-monotonic curve with maximized $k_v$ at intermediate $\alpha$ values. Together, the polygons and dashed lines outline the region covered by the parameter scan. Most experimental points fall within this region, indicating that the model captures the characteristic translational and rotational velocities. A few points outside this region are analyzed separately (SI Sec. 9). **Inset:** Effect of primary bending angle $\alpha$ on $k_v$ and $k_\Omega$. As $\alpha$ increases, $k_\Omega$ decreases monotonically. In contrast, $k_v$ reaches a maximum at $\alpha = 30° - 45°$.

## Supporting Information for

# Corkscrew motion of *Trypanosoma brucei* is driven by helical beating of the flagellum and facilitated by its bent shape

**Authors:** Sizhe Cheng[1], Devadyouti Das[1], Mykhaylo Barchuk[1], Raveen Armstrong[2], Michele M. Klingbeil[2], Becca Thomases[3], Shuang Zhou[1, *]

Corresponding author: zhou@physics.umass.edu

**This PDF file includes:**

Supporting text
Figures S1 to S11
Tables S1 to S2
Legends for Movies S1 to S4
SI References

**Other supporting materials for this manuscript include the following:**

Movies S1 to S4

**Supporting Information Text**

**1. Particle attachment and strength test**

The particle attachment is likely due to osmosis. We mix the *T. brucei* and fluorescent particles under different growth medium concentration and find a negative correlation between the number of attached particles and the growth medium concentration. We use optical tweezers experiments to verify the strength of attachment. While the posterior end of cells is captured by optical tweezers and dragged at speeds over $100\ \mu m \cdot s^{-1}$, the particles attached to their anterior surface remain in place, indicating a non-slip attachment.

**2. Calibration library and z determination**

We determine the z height of attached particles by comparing their defocused images with a pre-established height calibration library (Fig. S1B). To establish the library, the same fluorescent particles are attached to glass substrates, and defocusing images are taken as the microscope objective is moved at 0.5 μm increment. The center of the particle is tracked by ilastik-1.4.0 and set as the origin. By calculating azimuthally averaged intensity of the library image, we establish a look-up table of intensity ( $I_{lib}$ ) as a function of radius (*r*) at different height (*z*). The intensity of experimental data ( $I_{exp}$ ) is checked against the look-up table by computing the Pearson correlation coefficient

$$P_r = \frac{\sum_i (I_{exp}(r_i) - \overline{I_{exp}})(I_{lib}(r_i) - \overline{I_{lib}})}{\sqrt{\sum_i (I_{exp}(r_i) - \overline{I_{exp}})^2 \sum_i (I_{lib}(r_i) - \overline{I_{lib}})^2}} .$$

The position of the maximum $P_r$ determines the z-coordinates ( $z_0$ ) of experimental particles, as expressed by,

$$P_{rmax} = P_r(z_0)$$

The obtained $z_0(t)$ is a step wise trajectory because the step size of our calibration library is $0.5\ \mu m$. To increase the accuracy and smoothness, we perform an interpolation.

To understand how $P_r$ behaves between library points, we experimentally measure $P_r(\delta z)$ around a given library point, where $-0.5\ \mu m \le \delta z \le 0.5\ \mu m$ is the distance to that point. By finely tuning the objective position around the reference height, obtaining images at various $\delta z$ values, and calculating $P_r$ between these images and the reference image, we find that $P_r$ follows a quadratic relationship with $\delta z$ , as shown in Fig. S2B. This quadradic dependence is consistent across the height range of the library, but the exact curve shape, described by $P_r = a\delta z^2 + b$ , can slowly vary across the height. In practice, we do not perform fitting for each reference point to obtain the quadratic functions $P_r = a\delta z^2 + b$ . Instead, we find the coefficients $a$ and $b$ from the experimentally obtained distribution of $P_{rmax}$ . As shown in Fig. S2A, $P_{rmax}$ varies within a certain range due to imperfect alignment between the image and the reference points in calibration library. For the recorded period, particle height changes over a few micrometers, and the variation range of $P_{rmax}$ covers all $P_r$ values between the reference points. In this case, the maximum ( $P^+_{rmax}$ ) and minimum ( $P^-_{rmax}$ ) of $P_{rmax}$ represents special events: the point corresponding to $P^+_{rmax}$ should align with the reference point, while $P^-_{rmax}$ should lie midway between two adjacent reference points. Using these two points, $(0, P^+_{rmax})$ and $(step/2, P^-_{rmax})$ , we can obtain the quadratic interpolation function:

$$P_{rmax}(\delta z) = \frac{4(P_{rmax}^{-} - P_{rmax}^{+})}{step^2}\delta z^2 + P_{rmax}^{+}.$$

Therefore, after obtaining $P_{rmax}$ and $z_0$ by comparing the images, the offset ($\delta z$) of the image position relative to $z_0$ can be calculated from this quadratic function.

The next step is determining the sign of $\delta z$. We find the height $z_2$ of the second largest $P_r$. The actual height $z$ should fall between $z_0$ and $z_2$. Thus,

$$\text{sign}(\delta z) = \text{sign}(z_2 - z_0)$$

and,

$$z = z_0 + \delta z$$

Finally, due to the mismatch of the refractive index between air and the growth medium, these z-coordinate are rescaled by the ratio $n_{air} / n_{medium}$ to obtain the final *z* height (1). Combining the center of the rings with the final *z* height, we obtain the 3D position of the particle.

**3. Brownian motion verification**

We verify the reliability of our tracking method by measuring the diffusion coefficient of Brownian motion tracer particles. 20 individual fluorescent particles moving in the tape chamber are recorded by the PCO.edge camera. Each video is recorded at 20 fps for 15 seconds. The 3D particle tracking is applied, and the method of time lags is used to calculate the mean square displacement (MSD) in each direction.

$$MSD(t) = \frac{1}{T-t}\int_0^{T-t}[r(t'-t) - r(t')]^2 dt'$$

We use the Levenberg–Marquardt algorithm, which reduces to the ordinary least-squares method under linear conditions, to perform linear fitting through the Python package scipy.optimize.curve_fit and obtain the slope of each MSD plot (Fig. S3). We find that the slopes in the x, y, and z directions are highly consistent, with an average value of $1.596 \pm 0.049$ μm$^2 \cdot$ s$^{-1}$, and the overall three-dimensional slope is $4.789 \pm 0.026$ μm$^2 \cdot$ s$^{-1}$. The diffusion coefficient $D_{total} = Slope/(2n) = 0.798 \pm 0.004$ μm$^2 \cdot$ s$^{-1}$ is close to the diffusion coefficient $D_{est} = \frac{k_B T}{6\pi\eta r} = 0.795$ μm$^2 \cdot$ s$^{-1}$ estimated by the Einstein relation for 0.52 μm particles at 19 ℃, ensuring the reliability of this defocused particle tracking method.

**4. Definition of swimming direction**

To compare and analyze different trajectories, it is necessary to define a swimming direction and align all trajectories to move in the same orientation. Through defocusing particle tracking microscopy we get a 3D trajectory (Fig. S4A) that changes with time $x'(t), y'(t), z'(t)$. Using the least squares method, we calculate the slope $k_x, k_y, k_z$ and intercept $b_x, b_y, b_z$ of the regression lines in each direction by scipy.stats.linregress. We choose the 3D regression line $l'$ as the swimming direction,

$$l': \begin{cases} x_l' = k_x t + b_x \\ y_l' = k_y t + b_y \\ z_l' = k_z t + b_z \end{cases}.$$

After obtaining the forward direction $l'$, we translate the center of the trajectory to the origin and then rotate it around z-axis followed by the x-axis, so that the forward direction aligns with the positive z-axis. The transformed trajectory (Fig. S4B) is then expressed as $x(t), y(t), z(t)$.

**5. Cylindrical component extraction**

To obtain the cylindrical components of swimming trajectories (Fig. 2A), a coordinate transformation is required.

$$\begin{aligned}
r(t) &= \sqrt{x^2(t)+y^2(t)},\\
\theta(t) &= \arctan(y(t)/x(t)),\\
z(t) &= z(t).
\end{aligned}$$

Because $\arctan(y/x)$ is periodic and discontinuous at $\pm\pi/2$, we first use numpy.arctan2 to calculate the $\theta$ in the full range of $(-\pi,\pi)$ and then use numpy.unwrap to remove the jumps and reconstruct a smooth $\theta$ trajectory.

**6. Kinematic measurements on swimming cells**

From the changing rate and the oscillating frequency of $r(t)$, $z(t)$ and $\theta(t)$, we measured the beating frequency, velocity and angular velocity from 22 cells with attached particles (Fig. S5). $U$ and $|\Omega|$ measured from these swimmers generally increase as $f$ increases (Fig. S5D, E), but the dependence is too scattered to be described by a simple function, indicating that cell-to-cell differences can significantly contribute to the motility of *T. brucei*.

**7. Body shape measurements on active cells**

We measure the axial length $l$ (Fig. S6A) of the *T. brucei* (corresponding to $L$ in the simulations), as well as the radius $r$ (Fig. S6B) of the thickest part (corresponding to $D_3$ in the simulations) to ensure that our simulation parameters are reasonable. Based on 50 swimming cells, we find both values of the swimmers are relatively uniform, with $l = 19.6 \pm 1.7$ μm and $r = 1.43 \pm 0.16$ μm. From additional 22 cells attached to glass substrates, we also confirm that *T. brucei* is not straight and axisymmetric, but instead has 2 prominent junctions. The axes of the three body segments are approximated from the center line of time-lapse images (Fig. S6C) and the resulting angles are $\alpha = 42.8 \pm 11.4°$ and $\beta = 25.4 \pm 14.3°$, with their distributions shown in Fig. S6D and S6E.

**8. Dynamic deformation model and regularized Stokeslets simulation**

To fully understand the hydrodynamics consequence of flagellum motion, we simulate the locomotion of *T. brucei* using regularized Stokeslets method (2, 3). We first establish a body frame for the swimmer centered at $\boldsymbol{X}_0$, with its anterior end located at the origin. Then we define a body axis consisting of three line segments (Fig. 3), which correspond to the anterior section $L_1$, the middle section $L_2$ and the posterior section $L_3$. The angle between $L_1$ and $L_2$ is $\alpha$ and the angle between $L_2$ and $L_3$ is $\beta$. The three sections define a cell body plane ($\boldsymbol{e}_n$). $s$ is the position of surface points projected on the body axis. The beating amplitudes $A_n(s)$, $A_p(s)$ and body radius $r(s)$ at any position $s$ are obtained by the linear interpolation from four key positions: the anterior end, the first junction, the flagellum pocket (also second junction), and the posterior end (Table S2). To describe the dynamic changes of the body, we use transverse waves to represent the center line of the body $\boldsymbol{L}_c(s)$.

$$\boldsymbol{L}_c(s) = A_n(s)\sin(\omega t - ks)\boldsymbol{e}_n + A_p(s)\cos(\omega t - ks)\boldsymbol{e}_p(s) - \int_0^s \boldsymbol{e}_s(s')ds'$$

Here $\omega = 2\pi f$ and $k = 2\pi / \lambda$ are the input angular frequency and wave number. The wavelength of both oscillations is sent to be $\lambda = L/2 = 10\ \mu\text{m}$, which is estimated from our microscope images and suggested by other studies (4–9). $\boldsymbol{e}_p(s)$ (Fig. 3A) is in the body plane but perpendicular to the direction of local body axis (line segments) $\boldsymbol{e}_s(s)$.

For the regularized Stokeslets simulation, the cell body needs to be uniformly covered with $N$ point forces. At position $s$, a layer of surface points is placed along the circumference of a circle with center $\boldsymbol{L}_c(s)$, and radius $r(s)$, whose plane is perpendicular to $\boldsymbol{e}_s(s)$. The number of points $N_p(s)$ on such layer is determined by the input interval $ds$.

$$N_p(s) = \text{round}\left(\frac{2\pi r(s)}{ds}\right)$$

In our program, $ds = \frac{2\pi r(s)_{min}}{4} = 0.628\ \mu\text{m}$ is chosen to ensure that each layer contains at least four points. To identify each point on the circumference, we assign the $q$-th point a phase $\phi_q = q\frac{2\pi}{N_p(s)}, q = 1,2,...,N_p(s)$.

The total number of layers $N_l$ is estimated by keeping the distance between neighboring points in adjacent layers close to $ds$. If the body section is a frustum, $N_l$ is given by the rounded value of the slant height divided by $ds$. For example, in the anterior section, the number of layers is

$$N_{l1} = \text{round}\left(\frac{\sqrt{L_1^2 + \left(r(L_1) - r(0)\right)^2}}{ds}\right).$$

Then, the position of the center of the $m$-th circular layer $\boldsymbol{L}_c(s_m)$ can be obtained, where $s_m = m\frac{L_1}{N_{l1}}, m = 0,1,2,...,N_{l1}$.

Finally, the $i$-th surface point can be identified as the $q_i$-th point on the $m_i$-th circular layer, and its coordinates in the body frame can be written as,

$$\boldsymbol{X}_b^i = \boldsymbol{L}_c\left(s_{m_i}\right) + r\left(s_{m_i}\right)\left[\sin\left(\phi_{q_i}\right)\boldsymbol{e}_n\left(s_{m_i}\right) + \cos\left(\phi_{q_i}\right)\boldsymbol{e}_p\left(s_{m_i}\right)\right].$$

The corresponding velocity is estimated as,

$$\boldsymbol{u}_b^i = \frac{\boldsymbol{X}_b^i(t+dt) - \boldsymbol{X}_b^i(t)}{dt}.$$

Here, $dt = 1.25$ ms is the time step used in our simulation.

In the lab frame, the positions of these N points are:

$$\boldsymbol{X}_L^i(t) = \boldsymbol{R}\cdot\boldsymbol{X}_b^i(t) + \boldsymbol{X}_0,$$

where $\boldsymbol{R}$ is the rotation matrix connecting body and lab coordinates. The velocity of each point in the lab frame, $\boldsymbol{u}^i = \frac{d\boldsymbol{X}_L^i}{dt}$, can be written as:

$$\boldsymbol{u}^i = \boldsymbol{\Omega} \times \left( \boldsymbol{X}_L^i - \boldsymbol{X}_0 \right) + \boldsymbol{R} \cdot \boldsymbol{u}_b^i + \boldsymbol{u}_0 , \tag{1}$$

where $\boldsymbol{\Omega} = \left( \Omega_x, \Omega_y, \Omega_z \right)$ is the angular velocity and $\boldsymbol{u}_0$ is the translation velocity of the swimmer. The non-slip boundary condition requires that the flow velocity at the surface of the swimmer matches local motion, which is a superposition of the Green's function solution of Stokes equation, or Stokeslets $\boldsymbol{G}$:

$$\boldsymbol{u}(\boldsymbol{r}) = \int \boldsymbol{G}(\boldsymbol{r} - \boldsymbol{r}') \cdot \boldsymbol{f}(\boldsymbol{r}') d\boldsymbol{r}' ,$$

where $\boldsymbol{f}(\boldsymbol{r}')$ is the point force exerted at position $\boldsymbol{r}'$. We discretize the surface forces $\boldsymbol{f}(\boldsymbol{r}')$ into points and regularize them to small blobs, $\boldsymbol{F}_i(x) = \boldsymbol{f}_i \phi_\epsilon(\boldsymbol{x} - \boldsymbol{x}_0)$, where $\phi_\epsilon(\boldsymbol{x}) = \dfrac{15\epsilon^4}{8\pi\left(|\boldsymbol{x}|^2 + \epsilon^2\right)^{7/2}}$ is a popular form (10, 11) of radially symmetric function with $\int \phi_\epsilon(\boldsymbol{x}) d\boldsymbol{x} = 1$ and we choose $\epsilon = 1.4ds$ in our simulation. For the $N$ discrete point forces, $\boldsymbol{u}^i$ is then given by (11):

$$\boldsymbol{u}^i = \sum_{j=1}^{N} \frac{1}{8\pi\mu} \left( \frac{\left( |\boldsymbol{r}_{ij}|^2 + 2\epsilon^2 \right) \delta_{kl} + \left( \boldsymbol{r}_{ij} \right)_k \left( \boldsymbol{r}_{ij} \right)_l}{\left( |\boldsymbol{r}_{ij}|^2 + \epsilon^2 \right)^{3/2}} \right) \boldsymbol{f}_j = \sum_{j=1}^{N} M_{ij} \boldsymbol{f}_j , \tag{2}$$

where $\boldsymbol{r}_{ij} = \boldsymbol{X}_L^i - \boldsymbol{X}_L^j$ is the displacement between the $i$-th and $j$-th points. Free swimmers must also satisfy force-free and torque-free constraints, since no external force or torque is exerted on them:

$$\sum_{i=1}^{N} \boldsymbol{f}_i = 0 , \tag{3}$$

$$\sum_{i=1}^{N} \left( \boldsymbol{X}_L^i - \boldsymbol{X}_0 \right) \times \boldsymbol{f}_i = 0 . \tag{4}$$

By numerically solving equations (1)-(4), we determine $\boldsymbol{u}_0$, $\boldsymbol{\Omega}$ and $\boldsymbol{f}_i$ from the swimmer motion defined in the body frame, $\left( \boldsymbol{X}_b^i, \boldsymbol{u}_b^i \right)$. The trajectory of any point on the swimmer surface $\boldsymbol{X}_L^i$ can thus be calculated and compared with the experimental results (Fig. 1).

**9. Simulation parameters**

Although the attached fluorescent particles allow us to experimentally determine the beating frequency $f$ swimming speed $U$, angular velocity $\Omega$ and local beating amplitudes $A_n(s)$, $A_p(s)$ at the particle position $s$ for individual cells, they do not provide all biological details of a freely swimming *T. brucei*, such as $\alpha$, $\beta$, and the full $\left( A_n, A_p \right)$ profile. $\alpha$ and $\beta$ are measured from surface-attached cells, and $\left( A_n, A_p \right)$ profile is based on multiple particle positions across different cells (Fig. 2F). These quantities therefore represent typical values rather than one-to-one reconstruction of each free-swimming cell. The lack of means of direct comparison motivates us to perform a scan over the computational parameters and study the results in the motility space. In Fig. 5, we focus on examining how the primary bending angle $\alpha$ (between the anterior and middle section) and the beating amplitude at the anterior end influence the swimming characteristics. Therefore, we perform a parameter scan over $A_n(0)$, $A_p(0)$ and $\alpha$ while keeping other parameters fixed; the parameters we use are listed in Table S2.

The polygonal regions shown in Fig. 5 and Fig. S7 are constructed by connecting simulation points with the same $\alpha$ but varying amplitude, whereas the dashed curves connect points with the same amplitude but varying $\alpha$. Together, these scans delineate the region of motility space that can be generated by varying these two primary parameters. Although only $\alpha$ and the beating amplitude were varied in Fig. 5, most experimental points are within the region generated by this two-parameter scan.

Some experimental points appear outside the scan region because simulation results in Fig. 5 only shows a scan over three parameters, $A_n(0)$, $A_p(0)$ and $\alpha$. The full model allows variation of additional parameters of the *T. brucei* body, such as the $(A_n, A_p)$ profile and $\beta$. As an illustration, we present seven boundary points in Fig. S7 and identified their corresponding parameter sets, listed in Table S1. The comparison of experimental and simulation trajectories (points 1–5) are shown in Fig. 1.

**10. Discretization convergence and selection of the regularization parameter**

To test the convergence of the surface discretization, we varied the point interval $ds$ and computed the swimming speed and angular velocity, which are the two quantities of primary interest. Here, $c = 2\pi r(s)_{min}$ is the circumference of the smallest cross section. In our simulations, we chose $ds = c/4$, which ensures that each cross-sectional ring contains at least four discretization points. As shown in Fig. S9A, B, we kept the ratio $\epsilon / ds$ fixed and varied $ds$ from $c/2$ to $c/5$, which resulted in different total numbers of surface points $N_s$, shown on the horizontal axis. For each discretization, we calculated the mean normalized speed $V / V_{ds=c/4}$ and angular velocity $\Omega / \Omega_{ds=c/4}$ over all computational parameter sets (Table S2), with error bars indicating the range across these parameter sets. For all tested values of $\epsilon / ds$, both quantities approach stable values as the interval decreases, indicating convergence of the discretization. Specifically, for the value used in our simulations $\epsilon / ds = 1.4$ (black curve), decreasing the interval $ds$ from $c/4$ to $c/5$ changes the mean speed and angular velocity by less than 1% and 3%, respectively.

For the selection of $\epsilon$, it is generally taken to be close to $ds$, although there is no theoretical prescription for its exact value. We therefore performed a sensitivity test by fixing $ds$ and computing the speed and angular velocity for different values of $\epsilon / ds$. As shown in Fig. S9C,D, the results depend on $\epsilon$, but the region around $\epsilon / ds = 1.4$ shows the smallest rate of variation among the tested values, suggesting that $\epsilon / ds = 1.4$ is an appropriate choice for our simulation. Although varying $\epsilon$ changes the absolute values of the swimming speed and angular velocity, their relative dependence on the bending angle $\alpha$ remains unchanged. For all tested values of $\epsilon$, the angular velocity decreases with increasing $\alpha$, and at large beating amplitudes, the swimming speed consistently shows a maximum at an intermediate angle. This suggests that the existence of an optimal bending angle is intrinsic to the system rather than caused by numerical artifacts.

**11. Frustoconical sections as the minimal static model of cell shape**

At the anterior end of the flagellum, a short distal segment is free from the *T. brucei* body instead of being laterally attached. This segment is a natural extension of the flagellum and represents the starting point of the beating wave. In our experiments, chiral beating is observed along the entire cell body, so this behavior would naturally extend to this segment. Therefore, this free flagellar segment is not expected to change the body motion described in our model.

In terms of simulations, our model was constructed for the entire cell with an axial length of 20 μm, so the anterior region containing this free segment was already included in the model. We notice that the small diameter of this section (200-400nm) can lead to uncertainty of cell length measurement from images. However, their small volume and surface area are negligible compared with the entire cell. In addition, our minimal model was not intended to reconstruct fine surface details of the anterior end, which would require much higher geometric resolution and computational cost. Instead, approximating the anterior section of the cell as a frustum with gradually increasing radius captures the major structure feature and is consistent with our phase-contrast observations. In this sense, local surface details in this region may introduce small quantitative differences in the simulation results, but are not expected to have a major effect on the overall behavior.

## 12. Simulation of a hypothetically straight-body swimmer

We performed a straight-body simulation to show that it cannot reproduce the motion observed in experiments. We set $\alpha = \beta = 0$ to simulate the trajectory of the same point presented in Fig. 1B and Fig. 2, resembling a cell body that is straight and axisymmetric. From the trajectory, we see that the rotation radius $r_0 = 1.18$ μm of the marked point round the swimming direction $\mathbf{e}_z$ is much smaller (Fig. S8A) compared to a bent body (Fig. 2A). This is reasonable because as the bend shape is absent, the beating amplitudes determines the radius of the trajectory. The slope of $\theta(t)$ becomes positive, because the *ccw* rotation of the body is indistinguishable from the beating of the flagellum. The measured angular velocity, $\Omega' \approx 2\pi f + \Omega$, matches well with the slowed down angular velocity of *cw* beating velocity. Essentially, now what we see is the movement of the "petals" in the previous trajectory, but slowed down due to counter rotation. In the z-direction, the amplitude of oscillation is significantly reduced, because both oscillations of the helical beating are perpendicular to $\mathbf{e}_z$ due to axial symmetry. This is confirmed in the Fourier spectrum (Fig. S8B), where the peak at 7.28 Hz in the $z$ direction has been greatly reduced compared to the case of the real, bent-shape swimmer (Fig. 2B). These qualitatively deviations from experimental data provide strong evidence that the swimmer body can't be straight but should assume a bent shape.

## 13. Helical beating and body rotation lead to flower-like particle trajectories

The observed flower-like particle trajectories (Fig. 1 3D trajectory) belong to the class of hypotrochoidal curves. Such patterns arise naturally from the helical beating and body rotation in our model (Fig. S10). The trajectory of an attached particle under these two motions can be described as:

$$\begin{cases} x(t) = R_r\cos(\Omega t) + A_p\cos(\gamma)\cos(\Omega t)\sin(\omega t) + A_n\sin(\Omega t)\cos(\omega t) \\ y(t) = R_r\sin(\Omega t) + A_p\cos(\gamma)\sin(\Omega t)\sin(\omega t) - A_n\cos(\Omega t)\cos(\omega t) \\ z(t) = A_p\sin(\gamma)\sin(\omega t) + Ut \end{cases}$$

Here $\gamma$ is the angel between the axis of the particle-attached section and $\mathbf{e}_z$. $R_r$ is the body rotational radius, determined by both the body radius and the off-axis tilt, and represents the distance of the particle from the central axis. $\Omega < 0$ is the angular velocity of body rotation, while $\omega > 0$ is the angular frequency of chiral beating. $A_p\cos(\gamma)$ is the beating amplitude in the body plane and perpendicular to $\mathbf{e}_z$. $A_n$ is the beating amplitude normal to the body plane. If we assume that the chiral beating in the body frame traces a circle of radius $A$ in the transverse xy-plane, then we have $A_p\cos(\gamma) = A_n = A$. With this simplification, the xy-projection of the trajectory can be written as:

$$\begin{cases} x(t) = R_r\cos(\Omega t) + A\cos((\omega+\Omega)t - \frac{\pi}{2}) \\ y(t) = R_r\sin(\Omega t) + A\sin((\omega+\Omega)t - \frac{\pi}{2}) \end{cases}$$

Or, equivalently, in complex form:

$$W(t) = x + iy = R_r e^{i\Omega t} + Ae^{-i\frac{\pi}{2}}e^{i(\omega+\Omega)t}$$

With the substitutions $\varphi = \frac{\pi\Omega}{2\omega}$ and $\tau = -t + \frac{\pi}{2\omega}$, $W$ can be rewritten as

$$W(t) = e^{i\varphi}W_0(\tau)$$

where

$$W_0(\tau) = R_r e^{-i\Omega\tau} + Ae^{-i(\omega+\Omega)\tau}$$

The transformation from $W$ to $W_0$ only changes the orientation, starting point, and tracing direction of the curve. Thus, the two curves are geometrically equivalent. If we define $R = \frac{\omega}{\omega+\Omega} R_r$, $r = -\frac{\Omega}{\omega+\Omega} R_r$, $d = A$ and $\theta = -\Omega\tau$, then the real and imaginary parts of $W_0$ give the standard form of the hypotrochoid equation (12).

$$\begin{cases} x(t) = (R-r)\cos(\theta) + d\cos(\frac{R-r}{r}\theta) \\ y(t) = (R-r)\sin(\theta) - d\sin(\frac{R-r}{r}\theta) \end{cases}$$

The hypotrochoid is a special case of flower-like patterns, which occurs when $A_p\cos(\gamma) = A_n$. More generally, the beating parameters $A_p$ and $A_n$ control the shape of petals. When $A_p\cos(\gamma)$ is larger, the petals become longer and sharper (Fig. 1A), whereas when $A_n$ is larger, they become broader and flatter. If both $A_p$ and $A_n$ decrease gradually, the petals shrink and eventually disappear (Fig. 1C, D).

**14. Body-fixed simulation reveals the thrust and torque from helical beating**

To identify the force and torque that drive the translation and rotation of *T. brucei* during swimming, we designed a set of body-fixed simulations. Instead of requiring the force-free and torque-free conditions, we constrained the translational velocity $\boldsymbol{u}_0$ and angular velocity $\boldsymbol{\Omega}$ of the whole cell to be zero, while still allowing the cell to undergo helical beating. In this case, the velocity of each point in the body frame is identical to that in the lab frame $\boldsymbol{u}^i = \boldsymbol{u}_b^i$, and the corresponding force distribution is determined from Eq. 2.

$$\boldsymbol{u}^i = \boldsymbol{u}_b^i = \sum_{j=1}^{N} \frac{1}{8\pi\mu} \left( \frac{\left(\left|\boldsymbol{r}_{ij}\right|^2 + 2\epsilon^2\right)\delta_{kl} + \left(\boldsymbol{r}_{ij}\right)_k \left(\boldsymbol{r}_{ij}\right)_l}{\left(\left|\boldsymbol{r}_{ij}\right|^2 + \epsilon^2\right)^{3/2}} \right) \boldsymbol{f}_j = \sum_{j=1}^{N} M_{ij}\boldsymbol{f}_j$$

The total thrust force and reactive torque are given by,

$$\boldsymbol{F}_{th} = -\sum_{i=1}^{N} \boldsymbol{f}_i$$

$$\boldsymbol{T}_{ccw} = -\sum_{i=1}^{N} \left(\boldsymbol{X}_L^i - \boldsymbol{X}_c\right) \times \boldsymbol{f}_i$$

Here, we choose the body center $\boldsymbol{X}_c = \sum_{i=1}^{N} \boldsymbol{X}_L^i / N$ as the reference point for torque calculation, so that the torques obtained under different conditions can be compared consistently. Within the parameter range explored here (Table S2), the instantaneous thrust magnitude $\left|\boldsymbol{F}_{th}\right|$ is in the range of 0.5–1.7 pN, which is close to the propulsion forces previously measured for *T. brucei* with single developed flagella using optical trapping (13).

For long-term swimming, the more relevant quantities are the force and torque components along the swimming direction, since the transverse components cancel out when averaged over time. The swimming direction can always be identified from the free-swimming simulations (SI Sec. 8) and then aligned with $\boldsymbol{e}_z$ by rotation. Thus, these effective force and torque are obtained from the $\boldsymbol{e}_z$ components of the instantaneous force and torque, averaged over one beating period and can be written as $\boldsymbol{F}_{th}^b \cdot \boldsymbol{e}_z$ and $\boldsymbol{T}_{ccw}^b \cdot \boldsymbol{e}_z$. Figure S11C and D show the dependence of $\boldsymbol{F}_{th}^b \cdot \boldsymbol{e}_z$ and $\boldsymbol{T}_{ccw}^b \cdot \boldsymbol{e}_z$ on the bending angle $\alpha$ for $A_p(0) = A_n(0) = 3$ μm and $f = 7.2$ Hz. In particular, when $\alpha = 0$, $\boldsymbol{F}_{th}^b \cdot \boldsymbol{e}_z = 0.64$ pN and $\boldsymbol{T}_{ccw}^b \cdot \boldsymbol{e}_z = -14.41$ pN·μm.

**15. Simulations of body translation and rotation reveal the resistive force and torque**

To identify the resistive force and torque due to the translation and rotation of *T. brucei* during swimming, we performed a second set of simulations of body translation and rotation. In these simulations, we neither require the force-free and torque-free conditions nor allow the cell to beat. Instead, we take the cell shape at a given time and require the whole cell to translate with velocity $\boldsymbol{u}_0$ or rotate with angular velocity $\boldsymbol{\Omega}$. Therefore, the velocity of each surface point is directly determined: for the translation case, $\boldsymbol{u}^i = \boldsymbol{u}_0$, while for the rotation case, $\boldsymbol{u}^i = \boldsymbol{\Omega} \times (\boldsymbol{X}_L^i - \boldsymbol{X}_c)$. Using Eq. 2 again, the force distribution can be obtained, from which the resistive force and torque are calculated.

$$\boldsymbol{F}_t = -\sum_{i=1}^{N} \boldsymbol{f}_i^t \text{ and } \boldsymbol{T}_t = -\sum_{i=1}^{N} \left(\boldsymbol{X}_L^i - \boldsymbol{X}_c\right) \times \boldsymbol{f}_i^t$$

$$\boldsymbol{F}_r = -\sum_{i=1}^{N} \boldsymbol{f}_i^r \text{ and } \boldsymbol{T}_r = -\sum_{i=1}^{N} \left(\boldsymbol{X}_L^i - \boldsymbol{X}_c\right) \times \boldsymbol{f}_i^r$$

Here, the labels *t* and *r* denote the translation and rotation cases, respectively, corresponding to the resistive force and torque generated by cell translation and cell rotation. Since force and torque depend linearly on rigid-body translation and rotation at low Reynolds number, they can be related through a hydrodynamic resistance matrix.

$$\begin{pmatrix} \boldsymbol{F}_t + \boldsymbol{F}_r \\ \boldsymbol{T}_t + \boldsymbol{T}_r \end{pmatrix} = \begin{pmatrix} \overline{\overline{A}} & \overline{\overline{B}} \\ \overline{\overline{B}} & \overline{\overline{D}} \end{pmatrix} \begin{pmatrix} \boldsymbol{u}_0 \\ \boldsymbol{\Omega} \end{pmatrix}$$

The matrix elements of $\overline{\overline{A}}$, $\overline{\overline{B}}$, and $\overline{\overline{D}}$ in our model can be determined from the force and torque generated by unit translation or unit rotation along each coordinate direction.

To relate this matrix to long-time free swimming with average swimming speed $U\boldsymbol{e}_z$ and angular velocity $\Omega \boldsymbol{e}_z$, we consider an effective resistance matrix along the swimming direction, in which all matrix elements reduce to scalars.

$$\begin{pmatrix} F_t^\infty + F_r^\infty \\ T_t^\infty + T_r^\infty \end{pmatrix} = \begin{pmatrix} A & B \\ B & D \end{pmatrix} \begin{pmatrix} U \\ \Omega \end{pmatrix}$$

Here, the scalar force and torque represent the effective quantities along the swimming direction. Specifically, for each body shape over one beating period, we perform the corresponding translation or rotation simulation and extract the z-components of the resulting force and torque. The effective quantities are then obtained by averaging these z-components over the period.

$$F_t^\infty = \langle \boldsymbol{F}_t \cdot \boldsymbol{e}_z \rangle \text{ and } T_t^\infty = \langle \boldsymbol{T}_t \cdot \boldsymbol{e}_z \rangle$$

$$F_r^\infty = \langle \boldsymbol{F}_r \cdot \boldsymbol{e}_z \rangle \text{ and } T_r^\infty = \langle \boldsymbol{T}_r \cdot \boldsymbol{e}_z \rangle$$

For example, when $A_p(0) = A_n(0) = 3$ μm, $\alpha = 0$ and $f = 7.2$ Hz, the free-swimming simulation gives an average swimming speed of $U = 5.44$ μm/s and an average angular velocity of $\Omega = -18.09$ rad/s. Under this condition, the translational resistive force and torque are $F_t^\infty = -0.39$ pN and $T_t^\infty = 0.07$ pN·μm, while the rotational resistive force and torque are $F_r^\infty = -0.23$ pN and $T_r^\infty = 13.72$ pN·μm. In particular, $\frac{T_t^\infty}{T_r^\infty} \approx 0.5\%$ demonstrate that the translation of a helical body cannot provide enough torque to rotate it at the observed angular velocity, challenging the plane-rotational model.

**16. Force and torque balance at long-term limits**

Since the effective thrust force $\boldsymbol{F}_{th}^b \cdot \boldsymbol{e}_z$ (reactive torque $\boldsymbol{T}_{ccw}^b \cdot \boldsymbol{e}_z$ ) and the corresponding effective resistance $F_t^\infty$, $F_r^\infty$ ($T_t^\infty$, $T_r^\infty$) are obtained independently, the force and torque balance can be verified by summing the driving and resistive contributions. When $A_p(0) = A_n(0) = 3$ μm, $\alpha = 0$ and $f = 7.2$ Hz,

$$\boldsymbol{F}_{th}^b \cdot \boldsymbol{e}_z + F_t^\infty + F_r^\infty = 0.02 \text{ pN}$$

$$\boldsymbol{T}_{ccw}^b \cdot \boldsymbol{e}_z + T_t^\infty + T_r^\infty = -0.62 \text{ pN}\cdot\mu\text{m}.$$

These sums are very close to zero, with residuals of about 3% of $\boldsymbol{F}_{th}^b \cdot \boldsymbol{e}_z$ and 4% of $\boldsymbol{T}_{ccw}^b \cdot \boldsymbol{e}_z$. The small differences may arise from the averaging procedure used to construct the effective scalar quantities and only have a minor effect on the calculated force and torque.

By contrast, figure S11C, D show the dependence of $F_t^\infty$, $F_r^\infty$ and $T_t^\infty$, $T_r^\infty$ on the bending angle $\alpha$ for $A_p(0) = A_n(0) = 3$ μm and $f = 7.2$ Hz, with $U$ and $\Omega$ are determined from the force and torque balance rather than free-swimming simulation. The corresponding matrix elements are shown in Fig. S11B. This procedure satisfies the force and torque balance exactly, and the resulting force and torque differ slightly (~4%) from those calculated using the free-swimming $U$ and $\Omega$. This difference does not change their trends with $\alpha$ and thus its effect on the results of our analysis is negligible.

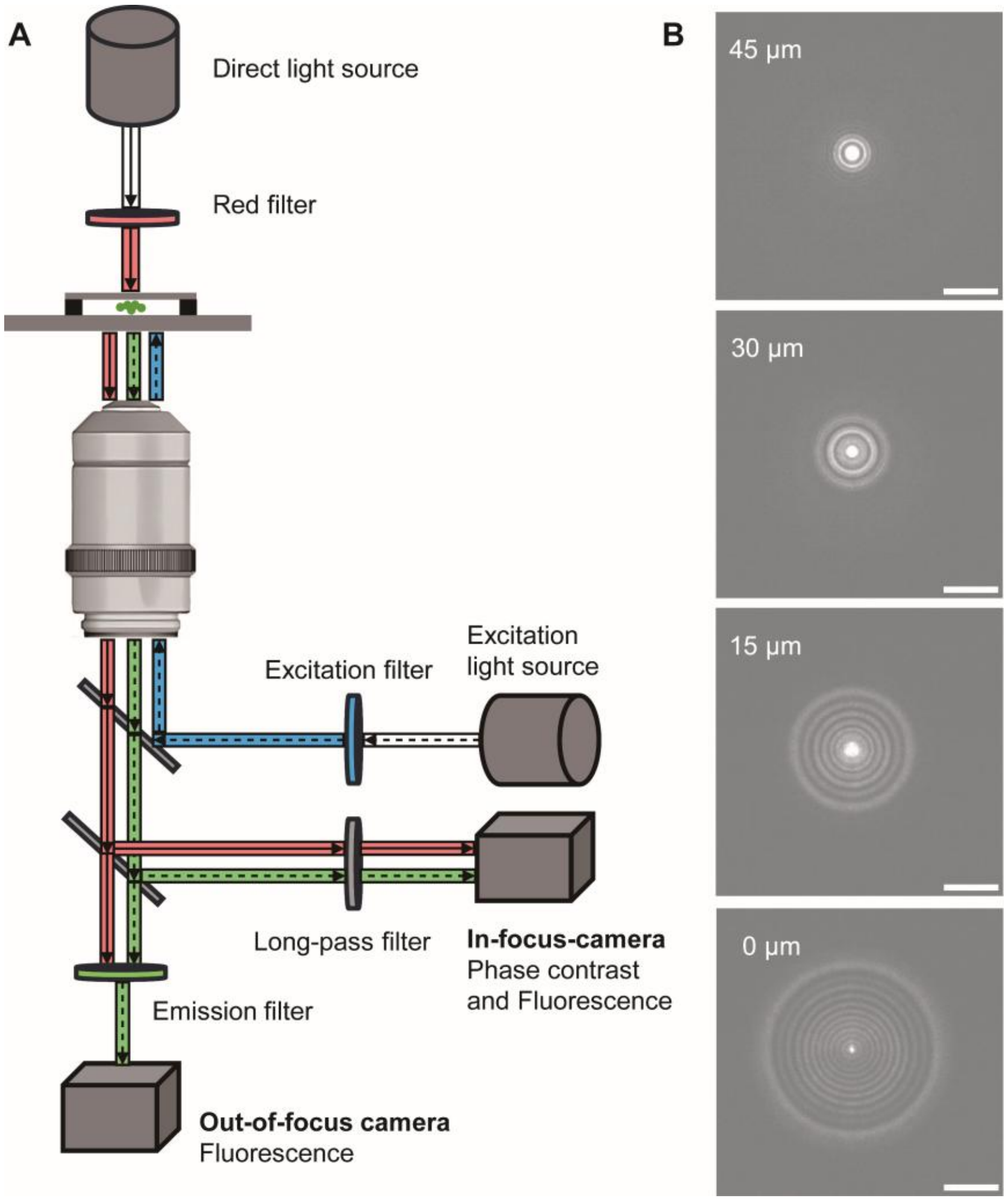

**Fig. S1. Experimental setup and height calibration library. (A)** Dual-channel defocusing particle tracking system. The sample is simultaneously illuminated by a transmitted red light ( $\lambda = 660$ nm ) and an epi-fluorescence excitation light ( $\lambda = 445$ nm ). The in-focus camera captures red phase-contrast images along with fluorescence signals through a long-pass filter (HQ505lp). The out-of-focus camera, placed behind an emission filter ( $\lambda = 510$ nm ), only records the defocused image of the fluorescent particle. **(B)** Defocused image library of a single fluorescent particle at different heights. A single fluorescent particle is fixed on a glass slide, while images are acquired by moving the objective in 0.5 μm steps. Results of every 15 μm are shown as an example. The numbers show the relative height of objective, with the starting point (the point farthest from the sample) defined as 0 μm . Scale bar: 10 μm .

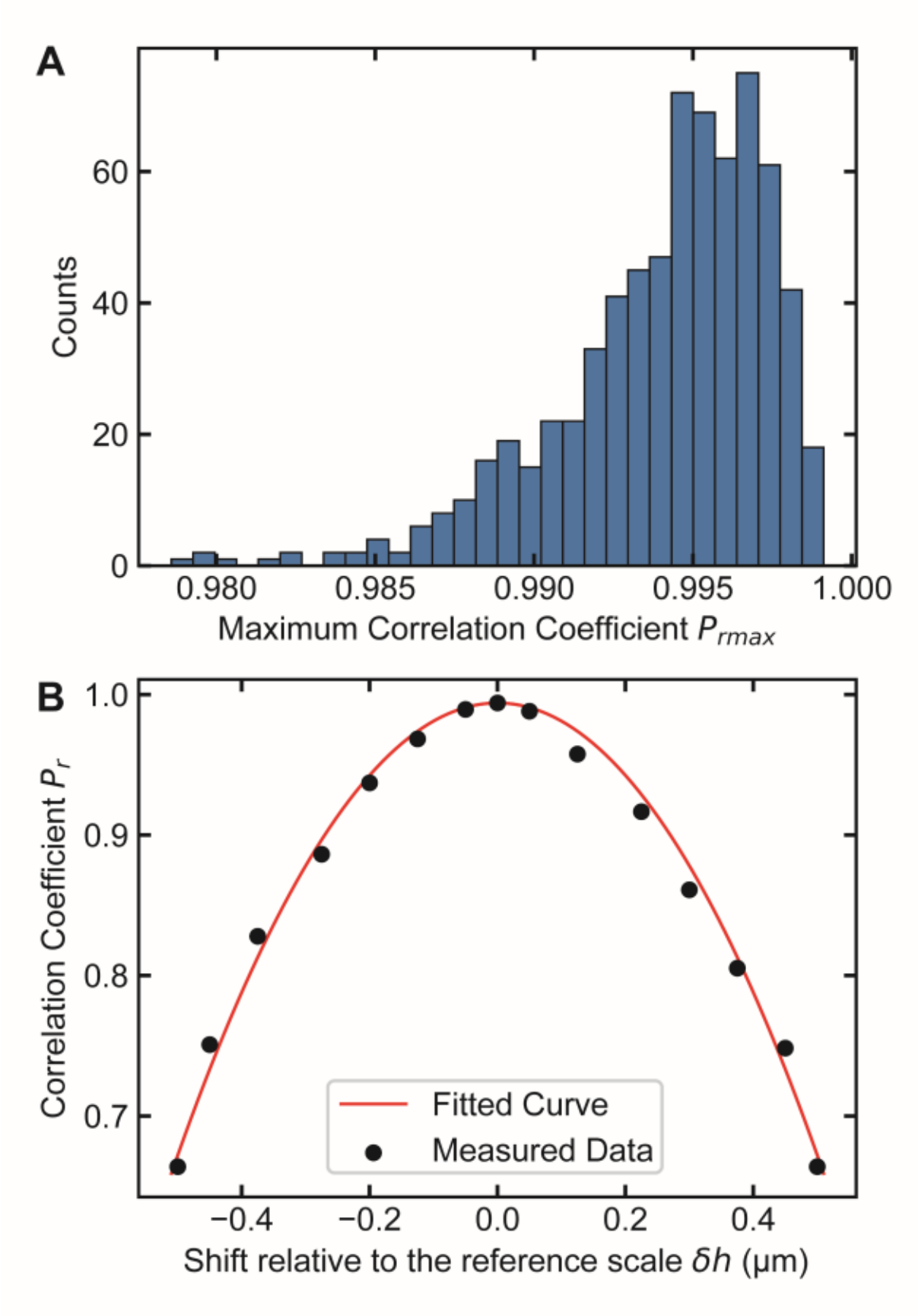

**Fig. S2. Interpolation based on the Pearson correlation coefficient. (A)** Histogram of maximum correlation coefficient $P_{rmax}$ obtained by comparing one particle image at anterior cell body (see Fig. 1B) with library images while *T. brucei* is swimming. **(B)** Correlation coefficient between the particle image and reference scale image as a function of the relative shift.

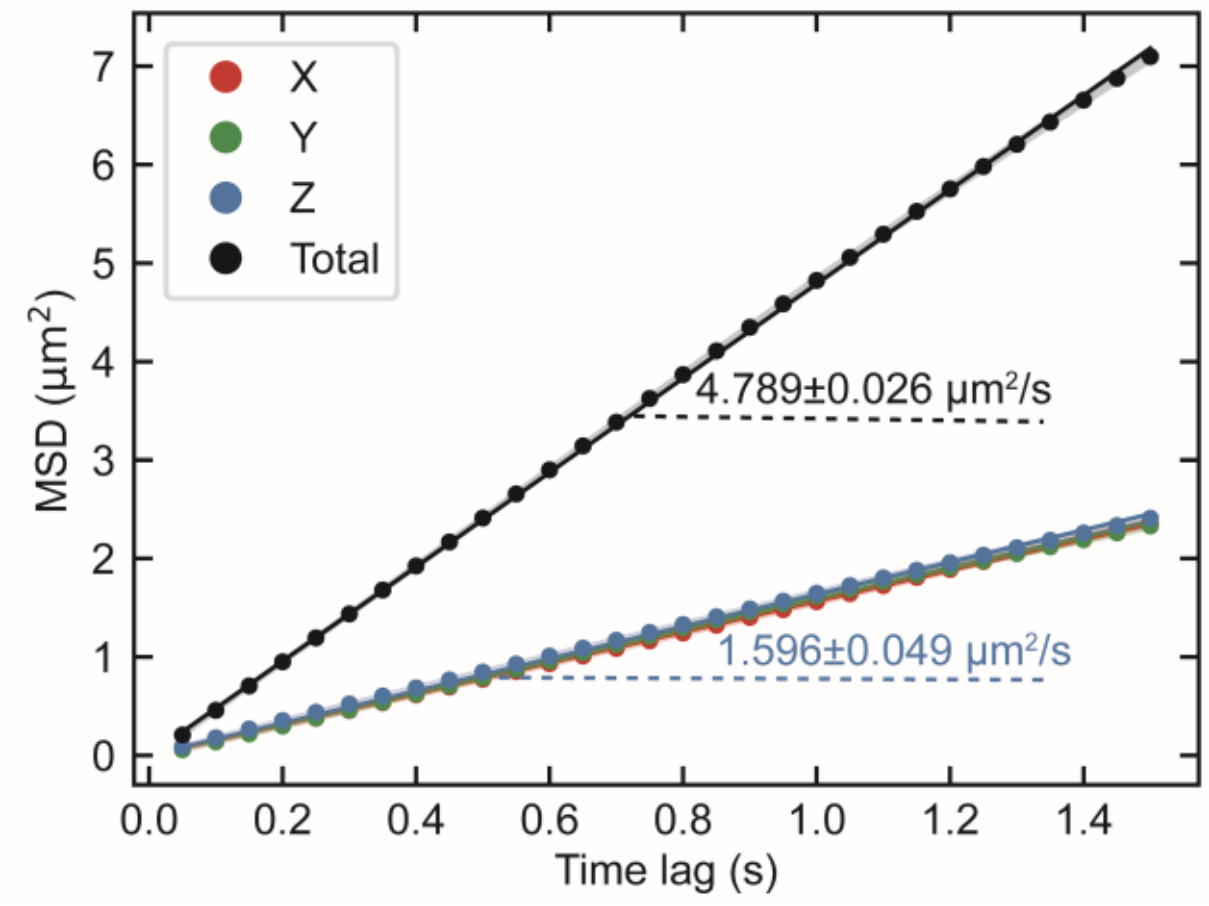

**Fig. S3. Measured mean squared displacement of 0.52 μm Brownian particles in water as a function of time.** The shaded area represents the standard error of the data points. For the total MSD, the number represents the slope obtained by linear fitting. For the MSDs of the *x*, *y*, and *z* components, the number represents the average slope across the three directions.

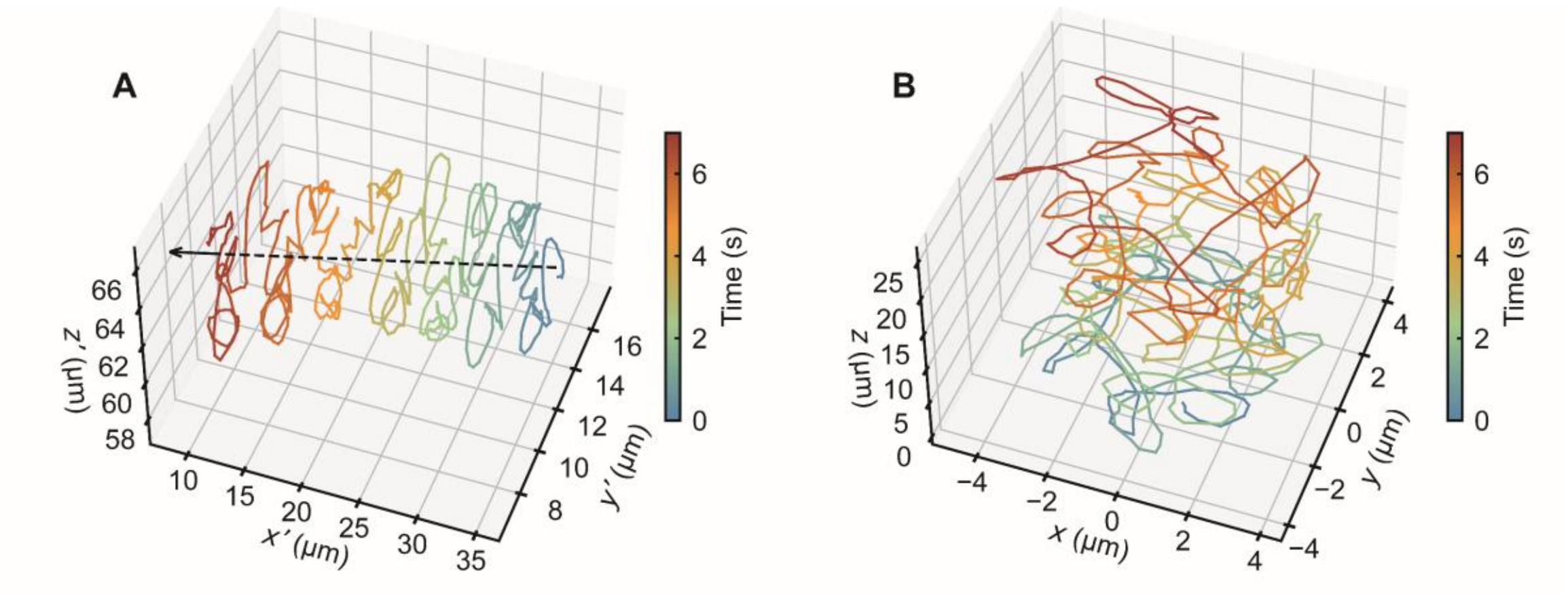

**Fig. S4. Coordinate transformation of trajectory. (A)** Original particle trajectory, with dashed line indicating the swimming direction $l'$. **(B)** Rotated particle trajectory, with the $+z$ direction as the swimming direction. The dynamics of this trajectory is shown in Movie S1.

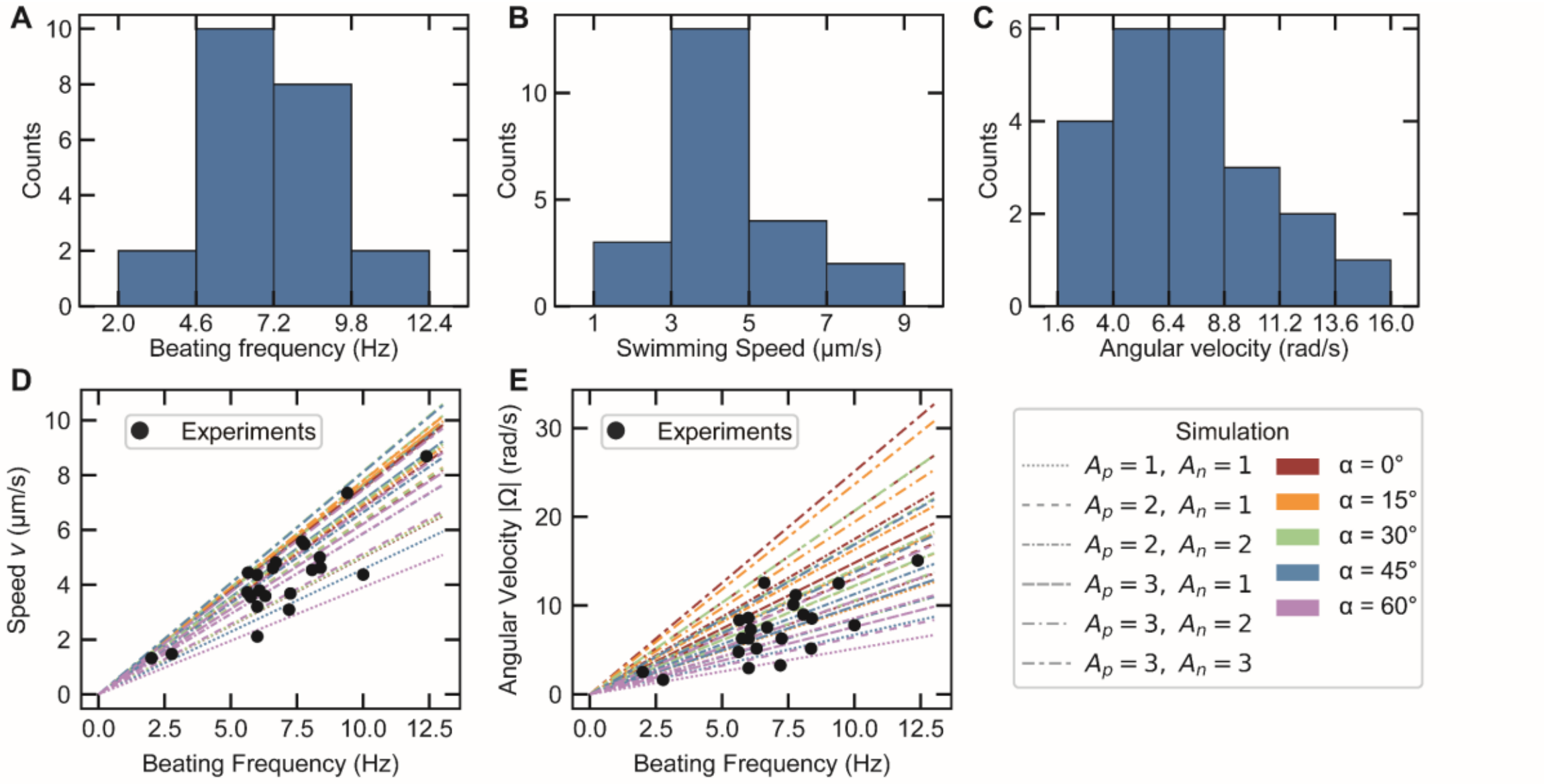

**Fig. S5. Statistics of the *T. brucei* motility.** Histogram of **(A)** beating frequency, **(B)** swimming speed and **(C)** angular velocity $|\Omega|$ measured from 22 particle labelled swimmers. **(D)** Swimming speed and **(E)** angular velocity versus beating frequency. Both quantities generally increase as beating frequency increases. Colored lines represent simulation results.

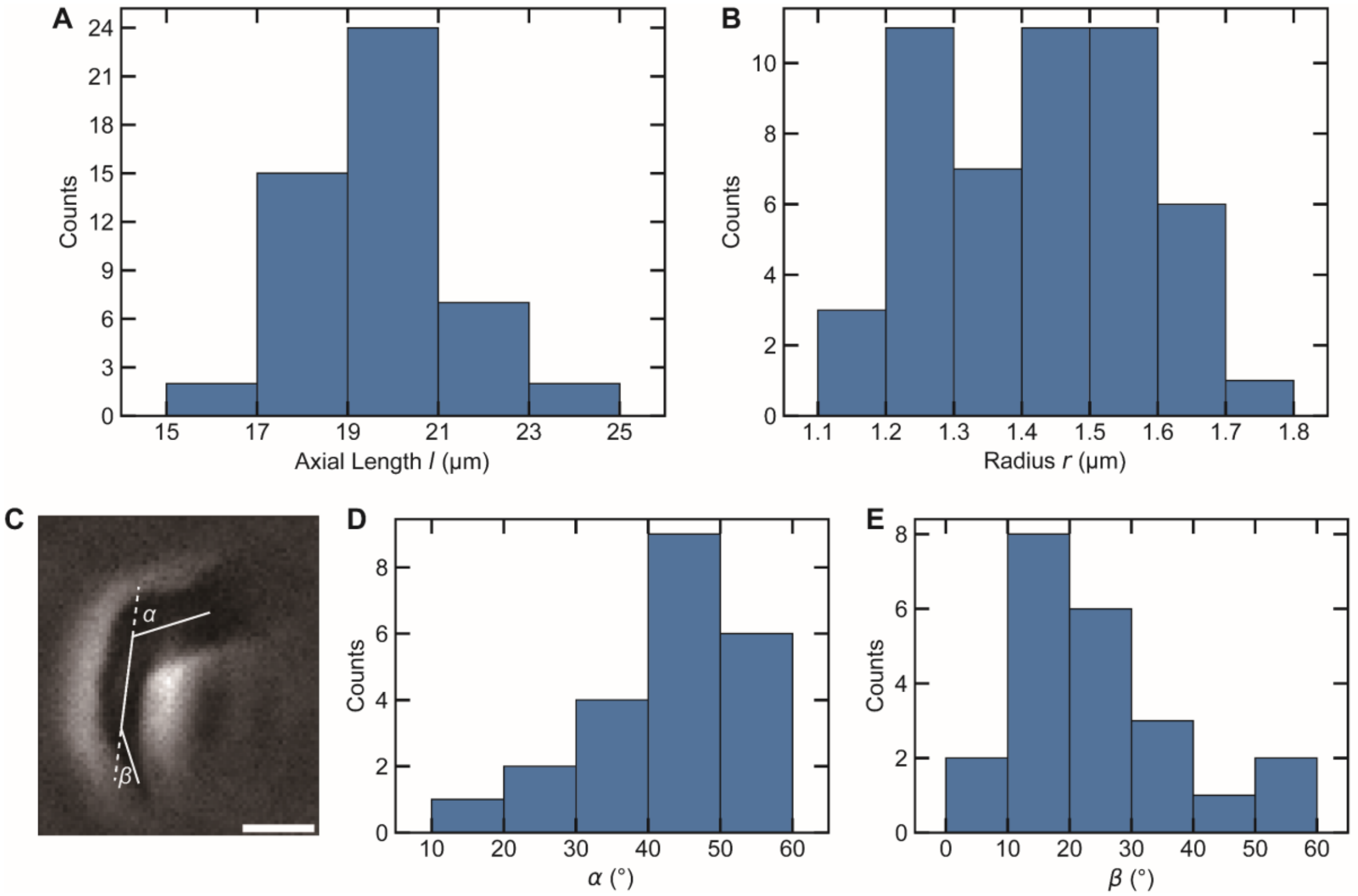

**Fig. S6. Body shape measurement of *T. brucei*. (A)** Histogram of body axial length $l$ measured from 50 swimming cells. **(B)** Histogram of the radius $r$ at the thickest part of the body, measured from 50 swimming cells. **(C)** Time-lapse image (minimum-intensity projection over 1.6 seconds) of single *T. brucei* stuck on the substrate and the estimated center line. Scale bar: 5 μm. **(D, E)** Histograms of bending angle $\alpha$ and $\beta$ measured from 22 cells stuck on the substrate.

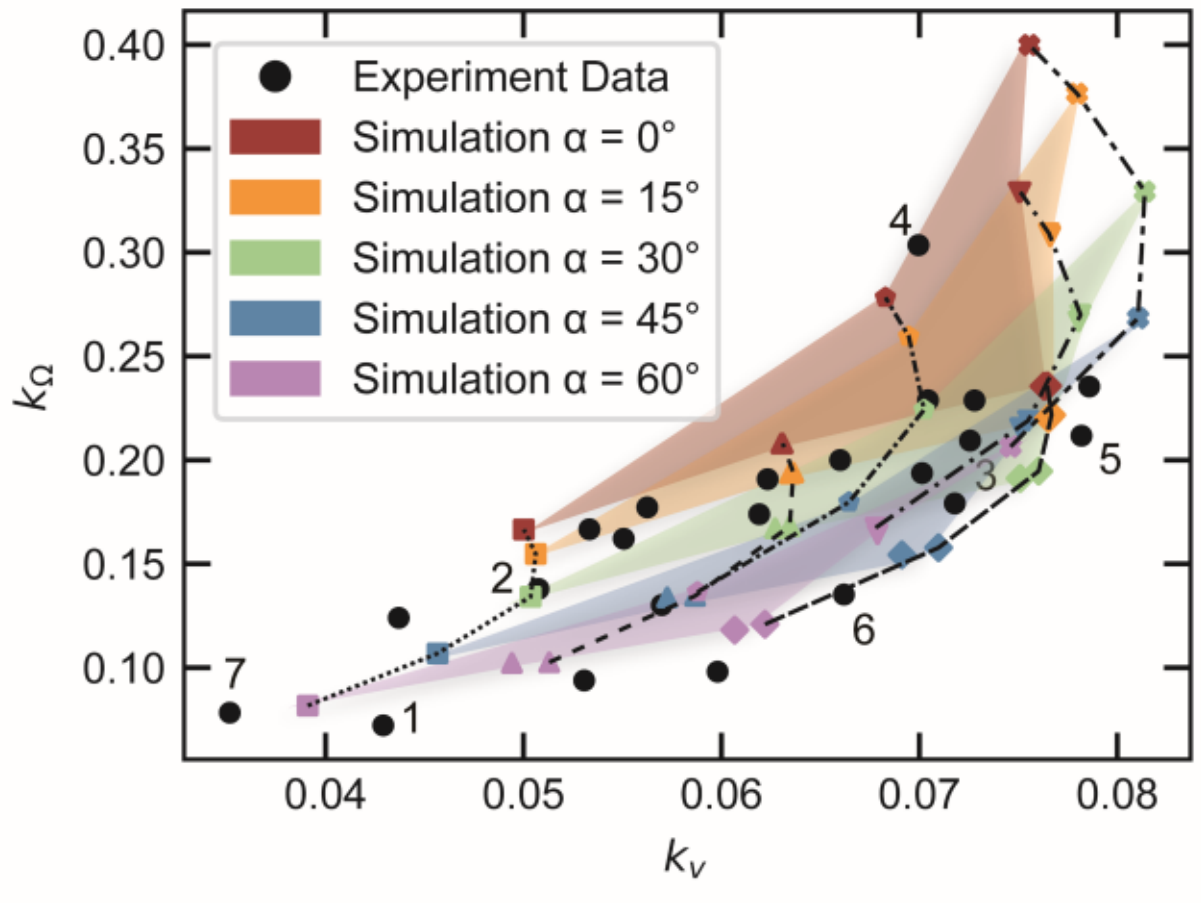

**Fig. S7. Experimental and simulated results in the motility space defined by the dimensionless translational and rotational velocities $k_v$ and $k_\Omega$.** The simulation parameters and their variation ranges are listed in Table S2. Numbers 1 to 7 represent the selected boundary points.

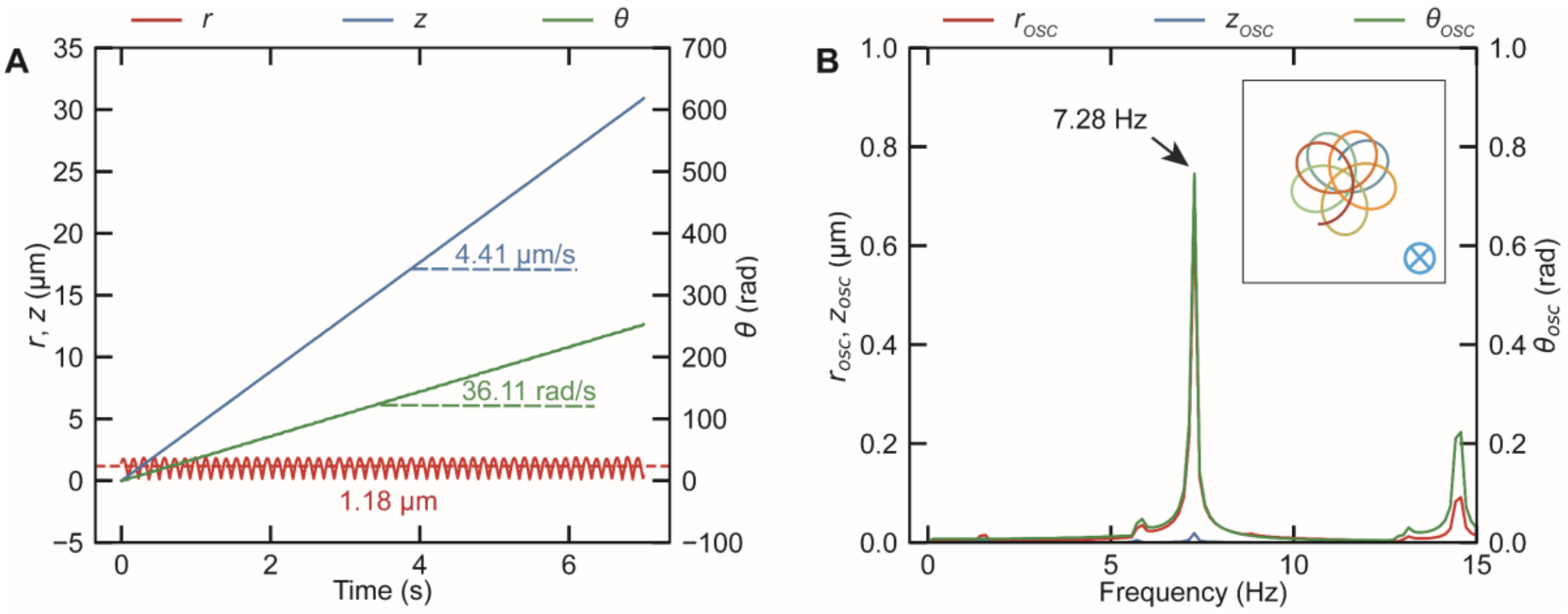

**Fig. S8. Temporal and Fourier analysis of the surface point on simulated straight body swimmer. (A)** Cylindrical coordinate components of the simulated trajectory at the same point as Fig. 1B from a simulation with a straight body shape. All beating amplitudes are set the same as Fig. 1B in simulation (see #2 in Table S1), but all bending angles are set to zero. The numbers represent the slopes obtained by linear fitting for the $z$ and $\theta$ trajectories, and the average radius for the $r$ component. **(B)** Fourier spectra of the simulated oscillating components after removing the linear part.

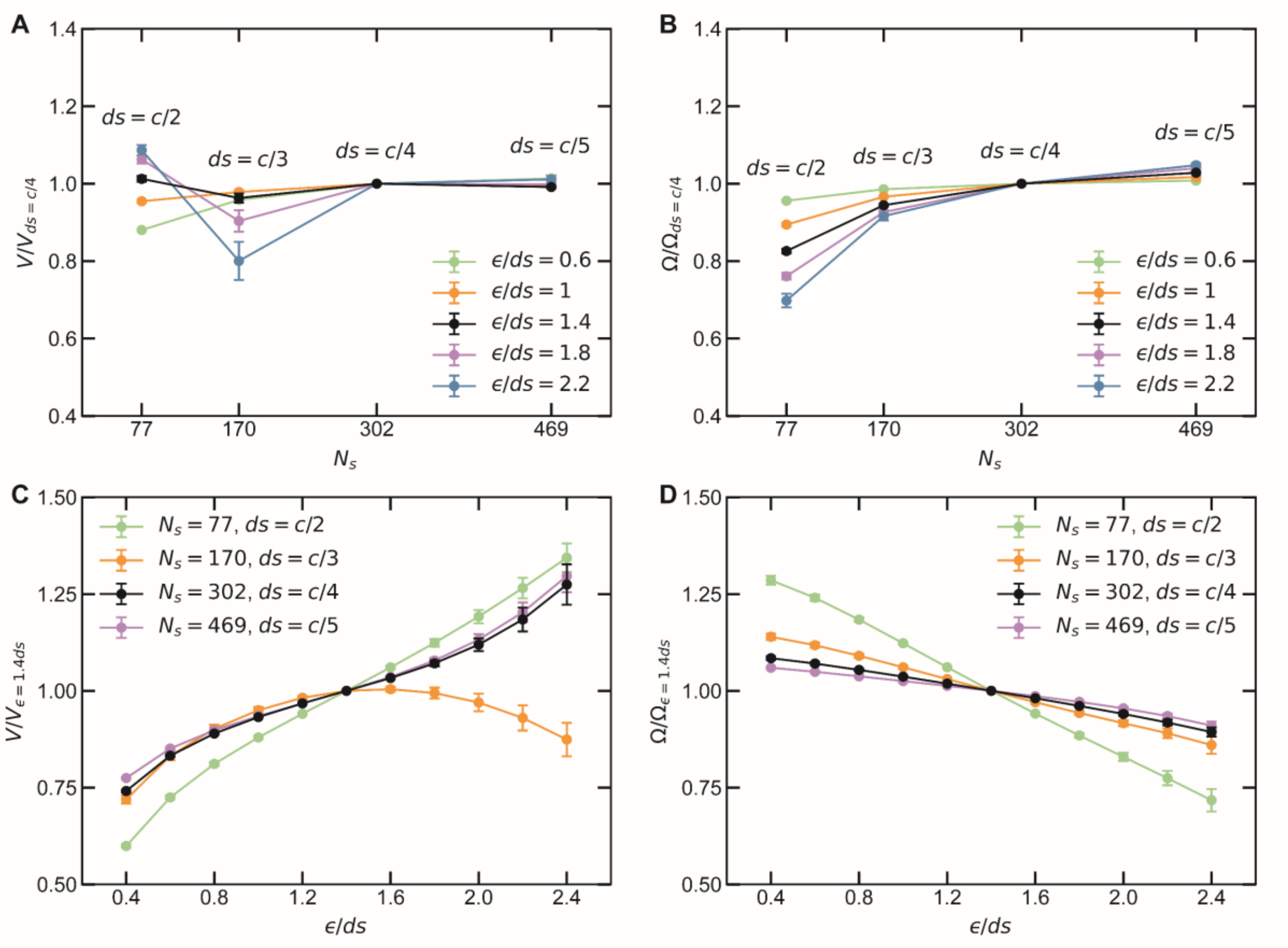

**Fig. S9. Numerical tests for the discretization spacing *ds* and regularization parameter $\varepsilon$ .** **(A, B)** Mean swimming velocity and angular velocity as functions of the number of discretization points $N_s$, with the corresponding spacing labeled as $ds = c / n$, where $c$ is the circumference of the smallest cross section in the model. Both quantities gradually approach stable values as $N_s$ increases, suggesting convergence with respect to the discretization. **(C, D)** Mean swimming velocity and angular velocity as functions of the relative regularization parameter $\epsilon / ds$ at fixed ds. The region around $\epsilon = 1.4ds$ shows the smallest rate of variation. The black curve represents the $\epsilon$ or $ds$ used in the current simulations, and the error bars indicate the range of variation across simulations with different beating parameters.

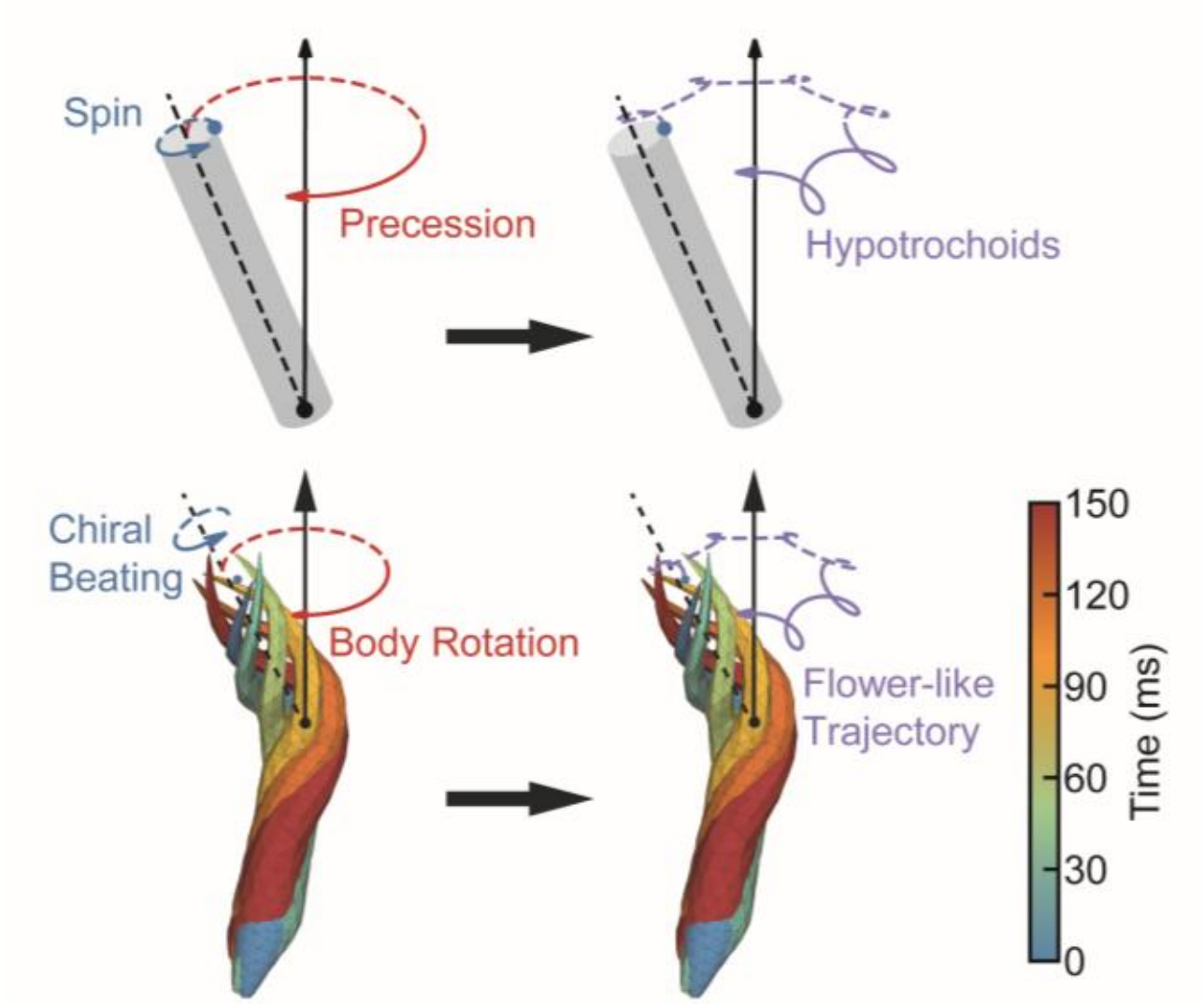

**Fig. S10. Geometric origin of flower-like trajectories. Top:** A hypotrochoid-like pattern generated by a point on a spinning cylindrical surface whose axis simultaneously precesses about one end. **Bottom**: A flower-like trajectory generated by chiral flagellar beating and slow body rotation in *T. brucei*.

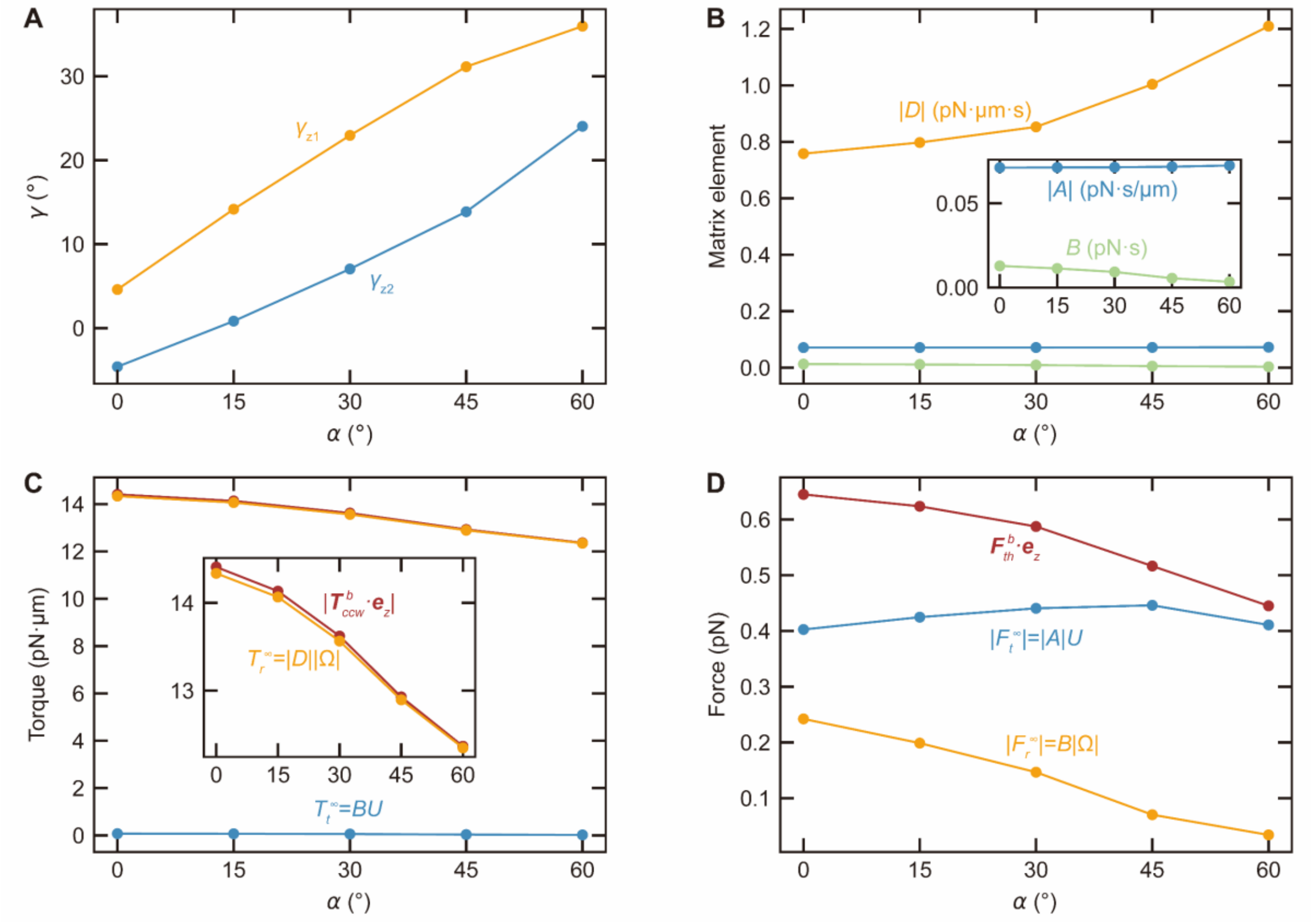

**Fig. S11. Effect of bent cell shape on swimming dynamics.** The simulations shown in this figure were performed using the parameters listed in Table S2, with $A_p(0)$ and $A_n(0)$ fixed at 3 μm. **(A)** As $\alpha$ increases, the angle between swimming direction and anterior section $\gamma_{z1}$ and the angle between swimming direction and middle section $\gamma_{z2}$ both increases. Their small nonzero values at $\alpha = 0$ arise from the nonzero $\beta$. **(B)** Dependence of the effective resistance matrix elements *A*, *B*, and *D* on the bending angle $\alpha$. **(C)** As $\alpha$ increases, both the reactive torque $|\boldsymbol{T}_{ccw}^{b} \cdot \boldsymbol{e}_z|$ and the resistance torque due to rotation $T_r^{\infty}$ decrease. The resistance torque due to translation $T_t^{\infty}$ is much smaller than the other two. **(D)** As $\alpha$ increases, both the thrust force $\boldsymbol{F}_{th}^{b} \cdot \boldsymbol{e}_z$ and the resistance force due to rotation $|F_r^{\infty}|$ decrease. The resistance force due to translation $|F_t^{\infty}|$ reach maximum around $\alpha = 45^{\circ}$.

| # | $\alpha$ (°) | $\beta$ (°) | Beating direction | Anterior end | First junction | Flagellum pocket | Posterior end |
|---|---|---|---|---|---|---|---|
| **1** | 30 | 15 | $A_p$ (μm) | 2.5 | 0.4 | 0 | 0.2 |
| | | | $A_n$ (μm) | 0.5 | 0.4 | 0 | 0.2 |
| **2** | 50 | 5 | $A_p$ (μm) | 1.4 | 0.8 | 0 | 0.7 |
| | | | $A_n$ (μm) | 1.4 | 0.8 | 0 | 0.7 |
| **3** | 45 | 15 | $A_p$ (μm) | 2.2 | 0.8 | 0 | 0.6 |
| | | | $A_n$ (μm) | 2.2 | 1 | 0 | 0.6 |
| **4** | 10 | 9 | $A_p$ (μm) | 2.8 | 0.6 | 0 | 0.6 |
| | | | $A_n$ (μm) | 2 | 1 | 0 | 0.8 |
| **5** | 10 | 9 | $A_p$ (μm) | 2.7 | 1 | 0 | 0.8 |
| | | | $A_n$ (μm) | 1.1 | 0.6 | 0 | 0.7 |
| **6** | 50 | 15 | $A_p$ (μm) | 3 | 0.8 | 0 | 0.6 |
| | | | $A_n$ (μm) | 1 | 0.8 | 0 | 0.6 |
| **7** | 60 | 15 | $A_p$ (μm) | 0.9 | 0.8 | 0 | 0.6 |
| | | | $A_n$ (μm) | 0.9 | 0.8 | 0 | 0.6 |

**Table S1. Computational parameters for fitting the experimental data near the boundary region (see Fig. S7).**

| Position | Anterior end | Anterior section | First junction | Middle section | Flagellum pocket | Posterior section | Posterior end |
|---|---|---|---|---|---|---|---|
| Angle (°) | | | 0~60 | | 15 | | |
| ***L*** (μm) | | 7 | | 9 | | 4 | |
| ***r*** (μm) | 0.40 | | 0.83 | | 1.50 | | 0.40 |
| $A_p$, $A_n$ (μm) | 1.0~3.0 | | 0.8 | | 0 | | 0.6 |

**Table S2. Default computational parameters used in the simulations of Fig. 5.**

**Movie S1. Combined movie of phase contrast, fluorescence, and particle tracking.** The four panels are synchronized in time, so that corresponding moments are shown simultaneously. **The Phase Contrast** view shows the position of the attached fluorescent particle on the cell body. The transparent *T. brucei* body hides the near–far relation and causes the direction of rotation to be ambiguous in this view. **The fluorescence** view reveals the actual particle height during motion through its defocus ring pattern. **Particle trajectory**, obtained from tracking the fluorescent particle, shows that the attached particle follows the chiral beating and unidirectional rotation of the cell body, forming a flower-like trajectory.

**Movie S2. Comparison between simulation and phase contrast view.**

**Movie S3. Points on the simulated body trace flower-like patterns and show chiral beating.** In the lab frame, the points trace flower-like trajectories, with large and sharp petals in the anterior region, smaller petals in the middle region, and no visible petals in the mid-posterior regions. In the *ccw* co-rotating frame, the anterior region shows clear *cw* chiral beating, whereas the mid-posterior region is nearly stationary. The swimming direction in this movie is along −z to better show the chiral beating.

**Movie S4. Bending shape of living cells stuck on the glass substrate.**

**SI References**